\documentclass[pra,aps,nopacs,twocolumn,twoside,superscriptaddress]{revtex4}

\usepackage{amsmath,amsfonts,amssymb,caption,color,epsfig,graphics,graphicx,hyperref,latexsym,mathrsfs,revsymb,theorem,url,verbatim,epstopdf,mathtools,enumerate}
\usepackage{subfigure}\usepackage{makecell}\usepackage{multirow}\usepackage{diagbox}
\hypersetup{colorlinks,linkcolor={blue},citecolor={blue},urlcolor={blue}}

\usepackage[qm]{qcircuit}

\newtheorem{definition}{Definition}
\newtheorem{proposition}[definition]{Proposition}
\newtheorem{lemma}[definition]{Lemma}

\newtheorem{theorem}[definition]{Theorem}
\newtheorem{corollary}[definition]{Corollary}
\newtheorem{conjecture}[definition]{Conjecture}

\newtheorem{remark}[definition]{Remark}
\newtheorem{example}[definition]{Example}
\newtheorem{question}[definition]{Question}
\newtheorem{memo}[definition]{Memo}

\def\squareforqed{\hbox{\rlap{$\sqcap$}$\sqcup$}}
\def\qed{\ifmmode\squareforqed\else{\unskip\nobreak\hfil
\penalty50\hskip1em\null\nobreak\hfil\squareforqed
\parfillskip=0pt\finalhyphendemerits=0\endgraf}\fi}
\def\endenv{\ifmmode\;\else{\unskip\nobreak\hfil
\penalty50\hskip1em\null\nobreak\hfil\;
\parfillskip=0pt\finalhyphendemerits=0\endgraf}\fi}
\newenvironment{proof}{\noindent \textbf{{Proof.~} }}{\qed}
\def\Dbar{\leavevmode\lower.6ex\hbox to 0pt
{\hskip-.23ex\accent"16\hss}D}
\makeatletter
\def\url@leostyle{%
  \@ifundefined{selectfont}{\def\UrlFont{\sf}}{\def\UrlFont{\small\ttfamily}}}
\makeatother
\def\bcj{\begin{conjecture}}
\def\ecj{\end{conjecture}}
\def\bcr{\begin{corollary}}
\def\ecr{\end{corollary}}
\def\bd{\begin{definition}}
\def\ed{\end{definition}}
\def\bea{\begin{eqnarray}}
\def\eea{\end{eqnarray}}
\def\beq{\begin{equation}}
\def\eeq{\end{equation}}
\def\bal{\begin{aligned}}
\def\eal{\end{aligned}}
\def\bem{\begin{enumerate}}
\def\eem{\end{enumerate}}
\def\bex{\begin{example}}
\def\eex{\end{example}}
\def\bim{\begin{itemize}}
\def\eim{\end{itemize}}
\def\bl{\begin{lemma}}
\def\el{\end{lemma}}
\def\bma{\begin{bmatrix}}
\def\ema{\end{bmatrix}}
\def\bpf{\begin{proof}}
\def\epf{\end{proof}}
\def\bpp{\begin{proposition}}
\def\epp{\end{proposition}}
\def\bqu{\begin{question}}
\def\equ{\end{question}}
\def\br{\begin{remark}}
\def\er{\end{remark}}
\def\bt{\begin{theorem}}
\def\et{\end{theorem}}
\def\bmm{\begin{memo}}
\def\emm{\end{memo}}

\def\btb{\begin{tabular}}
\def\etb{\end{tabular}}

\newcommand{\nc}{\newcommand}

\def\a{\alpha}
\def\b{\beta}
\def\g{\gamma}
\def\d{\delta}

\def\ve{\varepsilon}

\def\t{\theta}
\def\i{\iota}

\def\l{\lambda}

\def\x{\xi}
\def\p{\pi}
\def\r{\rho}
\def\s{\sigma}

\def\ps{\psi}
\def\og{\omega}

\def\L{\Lambda}

\nc{\bbA}{\mathbb{A}} \nc{\bbB}{\mathbb{B}} \nc{\bbC}{\mathbb{C}}
 \nc{\bbD}{\mathbb{D}} \nc{\bbE}{\mathbb{E}} \nc{\bbF}{\mathbb{F}}
 \nc{\bbG}{\mathbb{G}} \nc{\bbH}{\mathbb{H}} \nc{\bbI}{\mathbb{I}}
 \nc{\bbJ}{\mathbb{J}} \nc{\bbK}{\mathbb{K}} \nc{\bbL}{\mathbb{L}}
 \nc{\bbM}{\mathbb{M}} \nc{\bbN}{\mathbb{N}} \nc{\bbO}{\mathbb{O}}
 \nc{\bbP}{\mathbb{P}} \nc{\bbQ}{\mathbb{Q}} \nc{\bbR}{\mathbb{R}}
 \nc{\bbS}{\mathbb{S}} \nc{\bbT}{\mathbb{T}} \nc{\bbU}{\mathbb{U}}
 \nc{\bbV}{\mathbb{V}} \nc{\bbW}{\mathbb{W}} \nc{\bbX}{\mathbb{X}}
 \nc{\bbZ}{\mathbb{Z}}

 \nc{\bA}{{\bf A}} \nc{\bB}{{\bf B}} \nc{\bC}{{\bf C}}
 \nc{\bD}{{\bf D}} \nc{\bE}{{\bf E}} \nc{\bF}{{\bf F}}
 \nc{\bG}{{\bf G}} \nc{\bH}{{\bf H}} \nc{\bI}{{\bf I}}
 \nc{\bJ}{{\bf J}} \nc{\bK}{{\bf K}} \nc{\bL}{{\bf L}}
 \nc{\bM}{{\bf M}} \nc{\bN}{{\bf N}} \nc{\bO}{{\bf O}}
 \nc{\bP}{{\bf P}} \nc{\bQ}{{\bf Q}} \nc{\bR}{{\bf R}}
 \nc{\bS}{{\bf S}} \nc{\bT}{{\bf T}} \nc{\bU}{{\bf U}}
 \nc{\bV}{{\bf V}} \nc{\bW}{{\bf W}} \nc{\bX}{{\bf X}}
 \nc{\bZ}{{\bf Z}}

\nc{\cA}{{\cal A}} \nc{\cB}{{\cal B}} \nc{\cC}{{\cal C}}
\nc{\cD}{{\cal D}} \nc{\cE}{{\cal E}} \nc{\cF}{{\cal F}}
\nc{\cG}{{\cal G}} \nc{\cH}{{\cal H}} \nc{\cI}{{\cal I}}
\nc{\cJ}{{\cal J}} \nc{\cK}{{\cal K}} \nc{\cL}{{\cal L}}
\nc{\cM}{{\cal M}} \nc{\cN}{{\cal N}} \nc{\cO}{{\cal O}}
\nc{\cP}{{\cal P}} \nc{\cQ}{{\cal Q}} \nc{\cR}{{\cal R}}
\nc{\cS}{{\cal S}} \nc{\cT}{{\cal T}} \nc{\cU}{{\cal U}}
\nc{\cV}{{\cal V}} \nc{\cW}{{\cal W}} \nc{\cX}{{\cal X}}
\nc{\cZ}{{\cal Z}}

\nc{\hA}{{\hat{A}}} \nc{\hB}{{\hat{B}}} \nc{\hC}{{\hat{C}}}
\nc{\hD}{{\hat{D}}} \nc{\hE}{{\hat{E}}} \nc{\hF}{{\hat{F}}}
\nc{\hG}{{\hat{G}}} \nc{\hH}{{\hat{H}}} \nc{\hI}{{\hat{I}}}
\nc{\hJ}{{\hat{J}}} \nc{\hK}{{\hat{K}}} \nc{\hL}{{\hat{L}}}
\nc{\hM}{{\hat{M}}} \nc{\hN}{{\hat{N}}} \nc{\hO}{{\hat{O}}}
\nc{\hP}{{\hat{P}}} \nc{\hR}{{\hat{R}}} \nc{\hS}{{\hat{S}}}
\nc{\hT}{{\hat{T}}} \nc{\hU}{{\hat{U}}} \nc{\hV}{{\hat{V}}}
\nc{\hW}{{\hat{W}}} \nc{\hX}{{\hat{X}}} \nc{\hZ}{{\hat{Z}}}

\nc{\hn}{{\hat{n}}}

\def\diag{\mathop{\rm diag}}

\def\max{\mathop{\rm max}}
\def\min{\mathop{\rm min}}

\def\sch{\mathop{\rm Sch}}

\def\tr{\mathop{\rm Tr}}

\def\dg{\dagger}

\def\ox{\otimes}

\newcommand{\bra}[1]{\langle#1|}
\newcommand{\ket}[1]{|#1\rangle}
\newcommand{\proj}[1]{| #1\rangle\!\langle #1 |}

\newcommand{\norm}[1]{\lVert#1\rVert}
\newcommand{\abs}[1]{|#1|}

\def\Dbar{\leavevmode\lower.6ex\hbox to 0pt
{\hskip-.23ex\accent"16\hss}D}

\begin{document}


\title{Multipartite entangling power by von Neumann entropy}

\date{\today}
\author{Xinyu Qiu}\email[]{xinyuqiu@buaa.edu.cn}
\affiliation{LMIB(Beihang University), Ministry of Education, and School of Mathematical Sciences, Beihang University, Beijing 100191, China}
	\author{Zhiwei Song}\email[]{zhiweisong@buaa.edu.cn}
\affiliation{LMIB(Beihang University), Ministry of Education, and School of Mathematical Sciences, Beihang University, Beijing 100191, China}
\author{Lin Chen}\email[]{linchen@buaa.edu.cn (corresponding author)}
\affiliation{LMIB(Beihang University), Ministry of Education, and School of Mathematical Sciences, Beihang University, Beijing 100191, China}

\begin{abstract}
Quantifying the entanglement generation of a multipartite unitary operation is a key problem in quantum information processing. We introduce the definition of multipartite entangling, assisted entangling, and disentangling power, which is a natural generalization of the bipartite ones. We show that they are assumed at a specified quantum state. We analytically derive the entangling power of Schmidt-rank-two multi-qubit unitary operations by the minimal convex sum of modulo-one complex numbers. Besides we show the necessary and sufficient condition that the assisted entangling power of Schmidt-rank-two unitary operations reaches the maximum.  We further investigate the widely-used multi-qubit gates, for example, the entangling and assisted entangling power of the $n$-qubit Toffoli gate is one ebit. The entangling power of the three-qubit Fredkin gate is two ebits, and that of the four-qubit Fredkin gate is in two to $\log_25$ ebits.
\end{abstract}

\maketitle

\section{Introduction}

Entanglement is a key resource in quantum information processing tasks, such as dense coding \cite{bennett1992communication, qiu2022quantum}, distributed computation \cite{dis1999cirac} and efficient tomography \cite{cramer2010Efficient}.   In practice, entanglement is generated by nonlocal unitary operations.  It is of importance to quantify how much entanglement a unitary operation can generate. This is characterized by entangling power, the most fundamental quantity to evaluate the usefulness of the nonlocal unitary operation acting on a composed quantum system. For instance, more entanglement generated by a unitary operation guarantees high fidelity and robustness in quantum communication \cite{hu2021long, qiu2024w1}. 
Entangling power is firstly defined for the bipartite nonlocal unitary operation  \cite{ent2000zanardi, wang2002ep,hamma2004ep, mkz13}. 
By taking the maximum or average over all input states, the entangling power is defined  via various  entropy of entanglement, such as linear entropy \cite{ent2000zanardi} and von Neumann entropy \cite{Nielsen03}. Analytically, the entangling power of  Schmidt-rank-two bipartite unitary and some complex bipartite permutation unitaries are derived \cite{lcly20160808,ent2016yu,ent2018shen}.

Multipartite entanglement has been paid much attention with the development of its realization \cite{cao2023gene,high2023ever} and applications \cite{exp2022nad, expqkd2022nad}. Multipartite unitary operations are vital in the construction of such states. Their experimental implementation is paid much attention \cite{wang2015improving}, for example, the multi-qubit  controlled phase gate is implemented by extracting gate fidelity $\geq 97.4\%$ with neutral atoms \cite{levine2019parallel}. In this case, the investigation of entangling power of multipartite operations is necessary. 
 Multipartite entangling power with respect to the mean output entanglement for
 random pure states sampled according to the unitarily invariant Haar measure is investigated \cite{scott2004mul}.  
Recently it has been shown that the minimum entangling power is close to its maximum by choosing the entropy of entanglement or 
Schmidt rank as entanglement measure \cite{mini2019chen}.  The analytical expression of the entangling power  of  $n$-partite unitary operations is derived concerning  linear entropy of entanglement \cite{ent2020lin}, where the von Neumann entropy is not considered as it is much harder to obtain than the linear entropy. However, the von Neumann entropy is one of the cornerstones of quantum information theory \cite{ohya1995entropy, song2023proof}.  Roughly speaking, it quantifies the amount of quantum information contained in a state of independently distributed  quantum systems, and also fully characterizes single-shot state transitions in unitary quantum mechanics \cite{boes2019von}.
As an application, von Neumann entropy plays an essential role in the coding theorem \cite{schu1995coding,winter1999coding}. From a practical perspective, it is of importance to consider the multipartite entangling power by von Neumann entropy.  As far as we know, the analytical derivation of that has not been analyzed due to its difficulty.

In this paper, we focus on the property and analytical expression of entangling power of  $n$-partite unitary operations with respect to von Neumann entropy. To be specific, we show the definition of multipartite entangling, assisted entangling and disentangling entangling power in Definitions \ref{def:N-ep} and \ref{def:N-aep}.
It is the maximal von Neumann entropy over all input states with ancillae and possible bipartite divisions of multipartite systems. Hence the multipartite entangling power is a natural  generalization of the bipartite one. Based on that, we show that the multipartite entangling, assisted entangling and disentangling power are assumed at a specified state by some facts in functional analysis. They are presented in Propositions \ref{pro: aep max} and \ref{pro: ep max}. It implies that the supreme in the definition reaches the maximum exactly, and shows benefits in the derivation of entangling power. As the example, we show the analytical results of entangling power of three and $n$-qubit ($n\geq 4$) Schmidt-rank-two unitary operations  in Propositions \ref{pro:ep3qubit} and \ref{pro:ep-nqubit}, respectively. They are obtained by the results on the minimal convex sum of some modulo-one complex numbers in Lemma \ref{le: miniconv}.
  We  present the necessary and sufficient condition that the assisted entangling power of three and $n$-qubit ($n\geq 4$)  Schmidt-rank-two unitary operations reach the maximum in Propositions \ref{pro:aep-3qubit} and \ref{pro:aep-nqubit}, respectively.
Up to local unitaries, we consider the entangling and assisted entangling power of the widely-used multi-qubit unitary operations including Toffoli and Fredkin gate. We show that the entangling and assisted entangling power of a $n$-qubit Toffoli gate is one ebit in Proposition \ref{pro: n-toffoli ep}.  The entangling power of the three-qubit Fredkin gate is two ebits shown in Proposition \ref{pro: 3-fredkin ep}, and that of four-qubit Fredkin gate is in two to $\log_25$  ebits.

The rest of this paper is organized as follows. In Sec. \ref{sec:pre and not} we introduce the notations, some facts of bipartite entangling power, von Neumann entropy and functional analysis. In Sec. \ref{sec:def max=min} we show the definition and physical meaning of multipartite entangling power, and show their properties. In Sec. \ref{sec:ep mulqu sr2} we show the entangling and assisted entangling power of multi-qubit Schmidt-rank-two unitary operations. In Sec. \ref{sec:ep widely used}, we show the entangling and assisted entangling power of widely used gates including Toffoli and Fredkin gate. We conclude in Sec. \ref{sec:conclusion}.

\section{Preliminaries and notations}
\label{sec:pre and not}
In this section we show the notations and some facts used in
this paper. First we present the notations of this paper.
Then we show the definition and some properties of entangling power of bipartite unitaries. Finally we introduce some facts about continuous mapping in functional analysis, which will be used later.

Let $\cH_A$ be the Hilbert space of system $A$. We denote $\cM_{m,n}$ by the set of $m\times n$ complex matrices, $\cM_n$ by the set of $n\times n$ complex matrices and $\cU_n$ by the set of $n\times n$ unitary matrices, $\sch(U)$ by the Schmidt rank of a bipartite unitary.  The Schatten $p$-norm for arbitrary operator $A\in \cM_{m,n}$ and $p\in [1,+\infty)$ is defined as $\norm{A}_p=\big[\tr(A^\dg A)^{\frac{p}{2}}\big]^\frac{1}{p}$. By setting $p=1$ and $p=2$ in the definition of $\norm{\cdot}_p$, the trace norm and Frobenius norm are obtained, respectively. The distance induced by Schatten $p$-norm is given by $d_p(A,B)=\norm{A-B}_p$. We denote $E(\r)$ by the von Neumann entropy of the quantum state $\r$,
$K_E(U)$ by the entangling power of a $U\in \cU_n$, and $K_{\L:\L^c}(U)$ by the entangling power of $U$ under the bipartition $\L:\L^c$.

Next we present the definition of entangling and assisted entangling power of bipartite unitaries. 
The entangling power of a bipartite unitary $U$ acting on the Hilbert space $\cH$ of systems $A$,$B$ is defined as \cite{Nielsen03}
\begin{eqnarray}
    K_E(U)=\sup_{\ket{\a}\in \cH_{AR_A}, \; \ket{\b}\in \cH_{BR_B}}
    E(U(\ket{\a}\ket{\b})),
\end{eqnarray}
where $\ket{\a}$ and $\ket{\b}$ are pure states, $R_A$ and $R_B$ are the local auxiliary systems, and $E(\cdot)$ denotes the entanglement of the state $U(\ket{\a}\ket{\b})$, i.e. the von Neumann entropy $S(\cdot)$ of the reduced density matrix on any one system, $E(\ket{\ps_{AB}}):=S(\tr_A\proj{\ps})=S(\tr_B\proj{\ps})$.  No entanglement is required as the initial resource and $K_E(U)\geq 0$ for all bipartite unitaries $U$. From a generalized perspective, entanglement may be the initial resource. This derives the assisted entangling power, which is defined as \cite{lhl03}
\begin{eqnarray}
K_{E_a}(U)=\sup_{\ket{\ps}\in \cH_{AR_ABR_B}}
    [E(U(\ket{\ps}))-E(\ket{\ps})],  
\end{eqnarray}
where $\ket{\ps}$ is a bipartite pure state. Obvious it holds that $K_E(U)\leq K_{Ea}(U)$.
Any bipartite unitary $U$ acting on $\cH=\cH_A\otimes\cH_B$ has Schmidt rank equal to $n$ if there is an expansion of the form $U=\sum_{j=1}^nA_j\otimes B_j$, where $d_A\times d_A$ matrices $A_1,A_2,...,A_n$ are linearly independent, and the $d_B\times d_B$ matrices $B_1,B_2,...,B_n$ are also linearly independent. So $\sch(U)\leq \min\{d_A^2, d_B^2\}$. The operator Schmidt decomposition in a standard form, i.e. 
$U=\sum_{j=1}^rc_jA_j\otimes B_j$,
	where $\frac{1}{d_A}\tr(A_j^\dg A_k)=\frac{1}{d_B}\tr(B_j^\dg B_k)=\d_{jk}$, and the Schmidt coefficients satisfy $c_j>0$, $\sum_{j=1}^rc_j^2=1$. Obviously the Schmidt coefficients are invariant under local unitaries.
For any bipartite unitary $U$, the following inequality holds 
\begin{eqnarray}
\label{eq:Ksch<=KE}
	K_{\sch}(U)\leq K_E(U)\leq \log_2 \sch(U),
\end{eqnarray}
where  $K_{\sch}(U)=H(\{c_j^2\})=\sum_j-c_j^2\log c_j^2$ and $c_j$'s are defined in the standard form of Schmidt decomposition. If a bipartite unitary $U$ acting on $\cH_A\ox \cH_B$ is a controlled one, namely $U=\sum_{j=1}^m P_j\ox U_j$ with the pairwise orthogonal projectors $P_j$ on $\cH_A$ and the unitary operators $U_j$ on $\cH_B$, then the assisted entangling power of $U$ satisfies that 
\begin{eqnarray}
    \label{eq:aepU<=log2m}
K_E(U)\leq K_{E_a}(U)\leq \log_2m,
\end{eqnarray}
where the second inequality becomes equality if and only if there is a mixed state $\s\in \cH_B$ such that the equations $\tr(\s U_j^\dg U_k)=0$ hold for any $j,k$ and $j>k$ \cite{lcly20160808}. 
Using the above fact, one can obtain  assisted entangling power with the help of  entangling power in some cases.

In quantum information processing, functional analysis allows us to consider a specific analytical problem from the comprehensive perspective of pure algebra and topology. We introduce some facts about compact set of a Hilbert space and continuous mapping as follows.

\begin{theorem}
\label{th:fdns compact}
Let $H$ be a Hilbert space, and $Y$ is a subset of $H$. Then the subset $Y$ is compact if and only if $Y$ is closed and bounded in $H$.
\end{theorem}

\begin{definition} 
	\label{def: continuous map}
	Let $X=(X,d)$ and $Y=(Y,\bar{d})$ be metric spaces. A mapping $T: X\rightarrow Y$ is said to be continuous at a point $x_0\in X$ if for every $\ve>0$ there is a $\d>0$ such that $\bar{d}(Tx,Tx_0)<\ve$ for all $x$ satisfying $d(x,x_0)<\d$. $T$ is said to be continuous if it is continuous at every point of $X$.
\end{definition}
\begin{theorem}
	\label{cor:cm}
	A continuous mapping $T$ of a compact subset $M$ of a metric space $X$ into $\bbR$ assumes a maximum and minimum at some points of $M$.
\end{theorem}
Using above facts, we will show the essential property of entangling power. This property will show benefits in the analytical derivation of entangling power.

\section{ Definition and properties of multipartite entangling power}
\label{sec:def max=min}
In this section we propose the definition, physical meaning and some properties of multipartite entangling and assisted entangling power. In Sec. \ref{sec:def and men}, we show the definition and physical meaning of entangling and assisted entangling power of multi-qubit unitary operations. In Sec.  \ref{sec:sup=max} we show that the supremum is exactly maximum in their definitions, which will contribute to the calculation of the entangling power.
\subsection{Definition and physical meanings}
\label{sec:def and men}
 In this paper we focus on the multipartite quantum systems, where the structure of the entangled states is more complex than that of the bipartite case. For instance, an $n$-partite pure state $\ket{\ps}$ in the Hilbert space  $\cH=\cH_{A_1}\ox \cH_{A_2}\ox ...\ox \cH_{A_n}$ is fully separable in any bipartite cut if and only if $\ket{\ps}=\ox_{k=1}^n \ket{\ps_k}$ with $\ket{\ps_k}\in \cH_{A_k}$. However the violation of full separation may not lead to a fully entangled state, as a multipartite state may be entangled by the bipartition on some of the subsystems for example, the state $\ket{\g}_{A_1A_2\cdots A_n}=\ket{\a}_{A_1}\ox \ket{\b}_{A_2\cdots A_n}$, where $\a$ is a qubit state and $\ket{\b}$ is a $n-1$ qubit GHZ state. 

We study the entangling power of a multipartite unitary with regard to the maximal von Neumann entropy, over all possible divisions of a multipartite system. Its definition is given as follows.
\begin{definition}
\label{def:N-ep}
The entangling power of a unitary acting on the Hilbert space of a $n$-partite system $\cH=\cH_{A_1}\ox \cH_{A_2}\ox ...\ox \cH_{A_n}$  is defined as 
\begin{eqnarray}
&&K_E(U_{A_1,A_2,...,A_n}) \nonumber\\
&=&
\sup_{\ket{\ps_i}\in\cH_{A_iR_{i}}}\;\;
 \max_{\emptyset\subsetneq \L\subsetneq \{A_1R_{1},...,A_nR_{n}\}}
 \nonumber\\&& \quad
 \times E_{\L:\L^c}(U(\ket{\ps_1}_{A_1R_{1}}\ox...\ox\ket{\ps_n}_{A_nR_n}))
\nonumber	\\
&=&\max_{\emptyset\subsetneq \L\subsetneq \{A_1R_{1},...,A_nR_{n}\}}
 \;\;
\sup_{\ket{\ps_i}\in\cH_{A_iR_{i}}}
\nonumber\\&& \quad \times
E_{\L:\L^c}(U(\ket{\ps_1}_{A_1R_{1}}\ox...\ox\ket{\ps_n}_{A_nR_{n}})) ,
\nonumber
\end{eqnarray}
where $\L^c$ is the complement set of $\L$.
\end{definition}
One can see that the multipartite entangling power is a natural generalization of the bipartite one. Practically, if a gate $U$ is applied in a quantum protocol or circuit without initial entanglement, then the entangled generation of this protocol or circuit will not be larger than the entangling power of $U$. As mentioned above, the input states are chosen as fully separable states here. This means that we do not need any entanglement as the initial resource. The entangling power  $K_E(U_{A_1,A_2,...,A_n})$ shows the maximum entanglement that generated by $U_{A_1,A_2,...,A_n}$ over all kinds of bipartite cuts. As in Sec. \ref{sec:def and men}, it holds that $K_E(U_{A_1,A_2,...,A_n})\geq 0$. 

The multipartite assisted entangling power is also a generalization of the bipartite case. It allows the entanglement to be the initial resource and is characterized by the maximal von Neumann entropy over all possible bipartite cuts. The multipartite assisted entangling power is defined by the difference between the entanglement of output and input states.
\begin{definition}
\label{def:N-aep}
Let $\cH$ be the Hilbert space of a $n$-partite system, that is, $\cH=\cH_{A_1}\ox \cH_{A_2}\ox ...\ox \cH_{A_n}$.  
Then

(i) The assisted entangling power of a unitary acting on $\cH$ is defined as 
\begin{eqnarray}
\label{eq:kea}
	&&K_{E_a}(U_{A_1,A_2,...,A_n})
=
\max_{\emptyset\subsetneq \L\subsetneq \{A_1R_{1},...,A_nR_{n}\}}
\;\;
\\
&& \quad \times
\sup_{\ket{\ps}\in\cH}
\Big(E_{\L:\L^c}(U\ket{\ps})-E_{\L:\L^c}(\ket{\ps}) \Big).
\nonumber
\end{eqnarray}

(ii) The disentangling power of a unitary acting on $\cH$ is defined as 
\begin{eqnarray}
\label{eq:ked}
&&K_{E_d}(U_{A_1,A_2,...,A_n})
=
\max_{\emptyset\subsetneq \L\subsetneq \{A_1R_{1},...,A_nR_{n}\}}
\\
&& \quad \times
\sup_{\ket{\ps}\in\cH}
\Big(E_{\L:\L^c}(U^\dg\ket{\ps})-E_{\L:\L^c}(\ket{\ps}) \Big).
\nonumber
\end{eqnarray}
Here $\L^c$ in (\ref{eq:kea}) and (\ref{eq:ked}) is the complement set of $\L$.
\end{definition}
The multipartite assisted entangling power characterizes the power of a unitary operation that can change the entanglement of a quantum state. The input states of entangling power are included in that of assisted entangling power, and hence $K_{E}(U_{A_1,A_2,...,A_n})\leq K_{E_a}(U_{A_1,A_2,...,A_n})$. In practice, the multipartite assisted entangling power characterizes the maximal change of entanglement in a quantum protocol.

Note that the auxiliary systems are  necessary in Definition \ref{def:N-ep} and \ref{def:N-aep}. For example, the three-qubit unitary operation $U=\proj{000}+\ket{001}\bra{100}
+\ket{010}\bra{001}+\ket{011}\bra{101}
+\ket{100}\bra{010}+\ket{101}\bra{110}+\ket{110}\bra{011}+\proj{111}$ satisfies that $U\ket{a,b,c}_{ABC}=\ket{c,a,b}_{ABC}$. So it can not generate any entanglement from the fully separable states. By adding the auxiliary systems, one can see that the entangling power of the gate $U$ is 2 ebits by (\ref{eq:Ksch<=KE}).

\subsection{Entangling, assisted entangling and disentangling power are assumed at a quantum state}
\label{sec:sup=max}
The multipartite entangling, assisted entangling and disentangling power are defined by taking the supreme over all quantum states. One may wonder whether entangling and assisted entangling power can always be assumed at a specified input state, as it will facilitate their derivation. Here we prove that the answer is yes by some facts in functional analysis. This implies that the supreme in Definitions \ref{def:N-ep} and \ref{def:N-aep} are exactly maximum. Besides, this fact contributes to the analytical derivation and precise estimation of entangling power. The proof of the facts in this section is shown in Appendix \ref{apx: property}.

Before showing the main results, we introduce some basic facts about the set consisting of multipartite pure states and pure product states, respectively. 
\begin{lemma}
	\label{le: p-norm compact}
	Suppose $\cX_n$ is the set of $n$-partite operators given by $M=\sum_{j_1=1}^{d_1^2}...\sum_{j_{n}=1}^{d_{n}^2}
	c_{j_1j_2...j_n}
	 M_{j_1}^{(1)}\otimes...\otimes M_{j_{n}}^{(n)}$  with $c_{j_1j_2...j_n}\in\bbC$ and  $M_{j_s}^{(s)}\in \cM_{d_s}$, $s=1,2,...,n$. 
 The Schatten-$p$ norm is
 denoted by  $\norm{\cdot}_p$ with $p\in[1,+\infty]$. Then

(i)All density operators of $(d_1d_2...d_n)$-dimensional $n$-partite pure states form a compact subset of the normed space $(\cX_n,\norm{\cdot}_p)$.
 
(ii) All density operators of $(d_1d_2...d_n)$-dimensional $n$-partite pure product states form a compact subset of the normed space $(\cX_n,\norm{\cdot}_p)$.
\end{lemma}

\begin{remark}
 Lemma \ref{le: p-norm compact} also holds for the set of all quantum states, as the limit of a sequence of positive semidefinite matrices is also positive semidefinite. 
\end{remark}

Now we are ready to show the main result of this section. It implies that the supreme in the definition of assisted entangling and disentangling power is the maximum exactly.
\begin{proposition}
	\label{pro: aep max}
 (i) The supremum of multipartite assisted entangling power of unitary operations is assumed at some pure states. That is, the assisted entangling power  of a unitary operation $U$ acting on the $n$-partite system $\cH=\cH_{A_1}\ox \cH_{A_2}\ox ...\ox \cH_{A_n}$  is exactly  $K_{E_a}(U)
=
\max_{\emptyset\subsetneq \L\subsetneq \{A_1R_{1},...,A_nR_{n}\}}
\max_{\ket{\ps}\in\cH}
\big(E_{\L:\L^c}(U\ket{\ps})-E_{\L:\L^c}(\ket{\ps}) \big)$.
 
(ii) The supremum of multipartite disentangling power of unitary operations is assumed at some pure states. That is, the disentangling power of a unitary operation $U$ acting on the $n$-partite system $\cH=\cH_{A_1}\ox \cH_{A_2}\ox ...\ox \cH_{A_n}$ is exactly $K_d(U)=K_{E_a}(U^\dg)=
\max_{\emptyset\subsetneq \L\subsetneq \{A_1R_{1},...,A_nR_n\}}
\max_{\ket{\ps}\in\cH}
\big(E_{\L:\L^c}(U^\dg\ket{\ps})-E_{\L:\L^c}(\ket{\ps}) \big)$.
\end{proposition}
  
Similar to Proposition \ref{pro: aep max}, we obtain the following results concerning entangling power. 
 \begin{proposition}
\label{pro: ep max}
	 The maximum of multipartite entangling power of unitary operations is assumed at some  product states. That is, the  entangling power of a unitary operation $U$ acting on the $n$-partite system $\cH=\cH_{A_1}\ox \cH_{A_2}\ox ...\ox \cH_{A_n}$ is exactly  $K_{E}(U)
	 =
	 \max_{\emptyset\subsetneq \L\subsetneq \{A_1R_{1},...,A_nR_{n}\}}
	 \max_{\ket{\ps}\in\cH}
	 E_{\L:\L^c}(U\ket{\ps})$.
\end{proposition}

The preceding facts show that it suffices to consider the maximum for the derivation of the entangling power of unitary operations. Next we show some examples concerning multi-qubit Schmidt-rank-two unitary operations and some widely used multi-qubit unitary operations.

 \section{Entangling and assisted entangling power of multi-qubit Schmidt-rank-two unitaries }
  \label{sec:ep mulqu sr2}
  
 Schmidt-rank-two multi-qubit unitaries are widely used in quantum information processing, for example, the multi-qubit controlled phase gate. In this section we show the entangling power of this kind of unitaries. In Sec. \ref{sec:preparation}, we present some useful facts that will be used for the derivation of the entangling power. In Sec. \ref{sec:ep nqubit}, we show the entangling and assisted entangling power of multi-qubit Schmidt-rank-two unitary operations. 
 \subsection{Preliminary lemmas}
 \label{sec:preparation}
 We consider the minimum convex sum of finite module-one complex numbers. First we show the expression of the minimum convex sum of a set. Based on  that, we prove that the minimum convex sum decreases with the cardinality of this set. The proof of the facts in this section is shown in Appendix \ref{apx:pre lemma}. 
\begin{lemma}
	\label{le: miniconv}
The minimum convex sum of the set  $\cC=\{e^{i\t_j}\}_{j=1}^{n}$ with $\t_j\in [0,2\pi)$ and $\t_j\leq \t_{j+1}$ is derived by
  \begin{eqnarray}
	\label{eg: mincj}
	&& \!\!\!\!\!\!
 \min_{c_1,c_2,...,c_{n}\geq 0,\;\sum_jc_j=1} \bigg|\sum_je^{i\t_j}c_j\bigg|
 \\
	=
	&& \!\!\!\!\!\!
 \begin{cases}
		0, 
		\;\mbox{for} \;\;
		0\leq\t_j-\t_{j-1}\leq \pi,
		\t_{n}-\t_1\geq \pi,
		\; j=2,...,n,\\
		\min_{1\leq j\leq n} \abs{\cos(\frac{\t_j-\t_{j+1\mod n}}{2})},
		\;\mbox{otherwise}.
	\end{cases}	
 \nonumber
\end{eqnarray} 
\end{lemma}

From the proof of Lemma \ref{le: miniconv}, one can obtain that the minimum convex sum decreases with the cardinality of a set. This fact will be used to obtain the entangling power of multi-qubit unitary operations.
\begin{lemma}
\label{le:minC1C2}
Suppose two sets $\cC_1=\{e^{i\a_j}\}_{j=1}^{n}$ and $\cC_2=\{e^{i\b_k}\}_{k=1}^{m}$ satisfies $n\leq m$ and $\cC_1\subset \cC_2$. 
Then 

(i) The minimum convex sum of the set $\cC_1$ is not less than that of $\cC_2$, that is, 
\begin{eqnarray}
\label{eq:minc1c2}
\min_{\substack{c_1,c_2,...,c_{n}\geq 0,\\\sum_jc_j=1}} \bigg|\sum_je^{i\a_j}c_j\bigg| 
\geq
\min_{\substack{c_1,c_2,...,c_{m}\geq 0,\\\sum_kc_k=1}} \bigg|\sum_ke^{i\b_k}c_k\bigg|.
\end{eqnarray}

(ii) The maximum convex sum of the set $\cC_1$ is not more than that of $\cC_2$, that is, 
\begin{eqnarray}
\label{eq:minc1c2}
\max_{\substack{c_1,c_2,...,c_{n}\geq 0,\\\sum_jc_j=1}} \bigg|\sum_je^{i\a_j}c_j\bigg| 
\leq
\max_{\substack{c_1,c_2,...,c_{m}\geq 0,\\\sum_kc_k=1}} \bigg|\sum_ke^{i\b_k}c_k\bigg|.
\end{eqnarray}
\end{lemma}

We show the expression of  entangling power of a Schmidt-rank-two bipartite unitary acting on $d_A\times d_B$ states, which will be used later in the analysis of multi-qubit case. 
From Eq. (18) in \cite{lcly20160808},
it has been obtained that the entangling power of the unitary $U=P_1\otimes I_{d_B}+P_2\otimes \sum_{j=1}^n e^{i\t_j}\proj{j}$  with orthogonal projectors $P_1,P_2$ and real parameters $\t_j$ shows 
\begin{eqnarray}
	\label{eq: schu=2 keu}
	&& \!\!\!\!\!\!
 K_E(U)
 \\
=&& \!\!\!\!\!\!
\max_{\substack{c_1,c_2,...,c_{d_B}\geq 0,\\\sum_jc_j=1}}H\bigg(\frac{1-\abs{\sum_je^{i\t_j}c_j}}{2},\frac{1+\abs{\sum_je^{i\t_j}c_j}}{2}\bigg).
 \nonumber
\end{eqnarray}

 From Lemma \ref{le: miniconv} and  (\ref{eq: schu=2 keu}), the entangling power of Schmidt-rank-two bipartite unitaries can be obtained as follows.
\begin{lemma}
	\label{pro:schu=2}
	Suppose  $U=P_1\otimes I_{d_B}+P_2\otimes \sum_{j=1}^{d_B} e^{i\t_j}\proj{j}$ is a $d_A\times d_B$ controlled unitary with orthogonal projectors $P_1,P_2$ and real parameters $\t_j$. Without loss of generality, one can assume that $\t_1=0$ and $\t_j\leq\t_{j+1}<2\pi$ for $1\leq j \leq d_B$. Then $U$ has the entangling power
\begin{eqnarray}
&& \!\!\!\!\!\! K_E(U)
\nonumber\\
=&& \!\!\!\!\!\!\begin{cases}
H(\frac{1}{2},\frac{1}{2})=1, 
\;\mbox{for} \;\;
0\leq\t_j-\t_{j-1}\leq \pi,
\t_{d_B}\geq \pi, j\geq 2,\\
\max_{1\leq j \leq d_B} 
H(\frac{1+\cos(\a)}{2},
\frac{1-\cos(\a)}{2}),
\; \mbox{otherwise},
\end{cases}
\nonumber
\end{eqnarray}
where $\a=(\t_j-\t_{(j+1) \mod n})/2$.
\end{lemma}

Next we simplify the analysis of entangling power.
That is, the critical states for multipartite unitaries with Schmidt rank equal to two can be removed during its calculation. 
\begin{lemma}
	\label{le:triepmax}
Up to local unitary, we assume that a $n$-partite  unitary $U_{A_1,A_2,...,A_n}=\sum _{j=1}^m(P_j)_{A_i}\otimes (U_{j})_{A_i^c}$ is controlled from $A_i$ side, where the projectors $P_j$'s are pairwise orthogonal to each other and $U_j$'s are pairwise orthogonal $(n-1)$-partite unitaries.   Under the bipartition $A_i:A_i^c$, the maximal entanglement generation $K_{A_i:A_i^c}(U)$  can be derived by removing the ancilla $R_i$. That is,
\begin{eqnarray}
&& \!\!\!\!\!\! K_{A_i:A_i^c}(U)
\nonumber\\
= && \!\!\!\!\!\!
\max_{\ket{\ps_k}\in\cH_{A_kR_k}}E_{A_iR_i:(A_iR_i)^c}(U(\ket{\ps_1}\ket{\ps_2}...\ket{\ps_n}))
\nonumber\\
\label{eq:max a1b1c1}
=&& \!\!\!\!\!\!\!\!\!\!
\max_{\substack{\ket{\ps_i'}\in\cH_{A_i}, k\neq i\\ \ket{\ps_k}\in\cH_{A_kR_k}}}
E_{A_i:(A_iR_i)^c}
(U(\ket{\ps_1}\cdots\ket{\ps_i'}\cdots\ket{\ps_n}))
\\
\label{eq:maxSqj}
=&& \!\!\!\!\!\!\!\!\!\!
\max_{\substack{q_j\geq 0, 
	\sum_{j=1}^m q_j=1,\\ \ket{\ps_k}\in\cH_{A_kR_k}, k\neq i}}
S\bigg(	\sum_{j=1}^m q_j U_j
\big(\ox_{k\neq i} \proj{\ps_k}\big) U_j^\dg
\bigg).
\end{eqnarray}
\end{lemma}

From the proof of Lemma \ref{le:triepmax}, when a $N$-partite unitary is controlled from more than one subsystem, their corresponding ancilla systems can also be removed for the calculation of entangling power. 
\begin{lemma}
\label{le:m:nbipartition}
Let $\emptyset\subsetneq \L\subsetneq \{A_1R_{1},...,A_nR_{n}\}$ and $\l=\{i|A_iR_i\in \L\}$. Up to local unitary, we assume that a $n$-partite  unitary $U_{A_1,A_2,...,A_n}=\sum _{j=1}^m(P_j)_{\L}\otimes (U_{j})_{\L^c}$ is controlled from $\L$ side, where the projectors $P_j$'s are pairwise orthogonal to each other and $U_j$'s are pairwise orthogonal unitaries.   Under the bipartition $\L:\L^c$, the maximal entanglement generation $K_{\L:\L^c}(U)$  can be derived by removing the ancilla $R_\l$. That is,
\begin{eqnarray}
	&&K_{\L:\L^c}(U)
\nonumber \\
 =&& \!\!\!\!\!\!
\max_{\ket{\ps_k}\in\cH_{A_kR_k}}E_{\L:\L^c}(U(\ket{\ps_1}\ox\ket{\ps_2}\ox...\ox\ket{\ps_n}))
	\nonumber\\	=&& \!\!\!\!\!\!
\max_{\substack{\ket{\ps_i'}\in\cH_{A_i}, \ket{\ps_k}\in\cH_{A_kR_k},\\i\in \l, k\in \l^c}}
	E_{\{A_i\}:\L^c}
	(U(\ox_{i}\ket{\ps_i'}\ox_{k}\ket{\ps_k}))
\nonumber	\\
=&& \!\!\!\!\!\!
\max_{\substack{q_j\geq 0, \sum_{j=1}^m q_j=1,
\nonumber\\ \ket{\ps_k}\in\cH_{A_kR_k}, k\in \l^c}}
	S\bigg(	\sum_{j=1}^m q_j U_j
	\big(\ox_{k\in \l^c} \proj{\ps_k}\big) U_j^\dg
	\bigg).
 \nonumber
\end{eqnarray}
\end{lemma}

As we shall see, these facts bring convenience for the derivation of the entangling power of Schmidt-rank-two unitary operations. 

\subsection{Entangling and assisted entangling and assisted entangling power of multi-qubit Schmidt-rank-two unitaries}
\label{sec:ep nqubit}
First we consider the entangling power for the three-qubit case, which is a generalization of the two-qubit one. Using Lemmas \ref{le: miniconv} and \ref{le:triepmax}, we have the following results.
\begin{proposition}
\label{pro:ep3qubit}
	Up to system permutation, a Schmidt-rank-two three-qubit unitary  is a controlled unitary $U=(P_1)_A\otimes( U_1)_{BC}+(P_2)_A\otimes (U_2)_{BC}$ with orthogonal projectors $P_1,P_2$ and $U_1=\sum_{k=0}^{1}\sum_{t=0}^{1}e^{i\t_{k,t}}\proj{k,t},\; U_2=\sum_{k=0}^{1}\sum_{t=0}^{1}e^{i\og_{k,t}}\proj{k,t}$ with $\t_{k,t},\og_{k,t}\in [0,2\pi)$. Let 
	\begin{eqnarray}
	\cA_1=\{\og_{00}-\t_{00}, \og_{01}-\t_{01}, \og_{10}-\t_{10}, \og_{11}-\t_{11} \} 
	\!\!\!\! \mod 2\pi,
\nonumber\\
	\cA_2=\{\t_{10}-\t_{00}, \t_{11}-\t_{01}, \og_{10}-\og_{00}, \og_{11}-\og_{01} \}
\!\!\!\!	\mod 2\pi,
\nonumber	\\		
	\cA_3=\{\t_{01}-\t_{00}, \t_{11}-\t_{10}, \og_{01}-\og_{00}, \og_{11}-\og_{10} \}
\!\!\!\!	\mod 2\pi,
 \nonumber
	\end{eqnarray}
	and $a_{m}^{(j)}$ be the $m$-th smallest element in $\cA_j$.
	Then $U$ has the entangling power
	\begin{eqnarray}
&& \!\!\!\!\!\! K_{E}(U) \\
  = && \!\!\!\!\!\!
		\begin{cases}
			1, 
			\;\;\exists j\;s.t.\;
			0\leq a_m^{(j)}-a_{m-1}^{(j)}\leq \pi, \;
			a_4^{(j)}-a_1^{(j)}\geq \pi, 
			\\
		\max_{\substack{1\leq j \leq 3\\1\leq m \leq 4} } H\Big(\frac{1+\cos(\b)}{2},\frac{1-\cos(\b)}{2}\Big),\;\mbox{otherwise},
   \nonumber
		\end{cases}
	\end{eqnarray}
 where $m\in \{2,3,4\}$ and  $\b=a_m^{(j)}-a_{(m+1)\mod 4}^{(j)}$.
\end{proposition}
The proof of Proposition \ref{pro:ep3qubit} is given in Appendix \ref{apx: n-qubit ep}.

From Proposition \ref{pro:ep3qubit}, one can obtain the assisted entangling power of some Schmidt-rank-two three-qubit unitary, where  the assisted entangling power reaches the maximum.
\begin{proposition}
\label{pro:aep-3qubit}
By the assumption in Proposition \ref{pro:ep3qubit}, the assisted entangling power of   a Schmidt-rank-two three-qubit unitary reaches the maximum, i.e. one ebit if and only if  there is a $ j\in \{1,2,3\}$ such that $0\leq a_m^{(j)}-a_{m-1}^{(j)}\leq \pi, \;
			a_4^{(j)}-a_1^{(j)}\geq \pi$, for $m=2,3,4$.
\end{proposition}
\begin{proof}
The proof of Proposition \ref{pro:ep3qubit} shows that the unitary $U$ is controlled by two terms under any possible  bipartition. We consider the bipartition $\L:\L^c$, where $\L^c$ denotes the controlled system. Suppose  the unitaries $U_1$, $U_2$ on system  $\L^c$ correspond to the set $\cA_j$ in Proposition \ref{pro:ep3qubit}.
From (\ref{eq:aepU<=log2m}), one has $K_{E_a}(U)\leq 1$ (ebit), and the entanglement generation by this bipartition $K_{\L:\L^c}(U)=1$ if and only if there is a mixed state $\s\in \cH_{\L^c}$ such that $\tr(\s U_1^\dg U_2))=0$, where $U_1,U_2$ is the controlled unitaries implemented on $\cH_{\L^c}$. Again by the proof of Proposition \ref{pro:ep3qubit},  $\exists \s\in \cH_{\L^c}$ s.t. $\tr(\s U_1^\dg U_2))=0$ if and only if the minimum convex sum of the set $A_j$ contains the original point, namely  $0\leq a_m^{(j)}-a_{m-1}^{(j)}\leq \pi, \;
			a_4^{(j)}-a_1^{(j)}\geq \pi, \; m=2,3,4$.
Note that $K_E(U)$ is the maximal  entanglement generation over all divisions. Hence $K_E(U)=1$ if and only if there is a $j\in \{1,2,3\}$ satisfies the condition above.
\end{proof}

Proposition \ref{pro:aep-3qubit} gives the necessary and sufficient condition that the three-qubit assisted entangling power reaches the maximum. It shows that many widely used three-qubit universal quantum gates have the assisted entangling power equal to one ebit, as we shall see in Sec. \ref{sec:ep widely used}. From the perspective of entanglement, these universal gates generate the maximum entanglement among all Schmidt-rank-two unitaries and hence are economic in quantum protocols.

Next we focus on the entangling and assisted entangling power of $n$-qubit ($n\geq 4$) Schmidt-rank-two unitaries.  The multipartite unitary gates are called genuine if they are not product unitary operators across any bipartition. 
In \cite{shen2022class}, the Schmidt-rank-two genuine multipartite unitary gates are classified, and the parametric Schmidt decomposition is given in TABLE \ref{tab:nqubit}.

\begin{table*}
\caption{The classification of genuine $n$-qubit ($n\geq 4$) unitary gates of Schmidt-rank-two under local equivalence}
\centering
\begin{tabular}{|c|c|c|}
\hline
singular number & parametric Schmidt decomposition & range of parameters  \\
\hline
$k=n$ & $I_2^{\otimes n}+(e^{i\phi}-1)\proj{0}^{\otimes n}$ & $\phi\in(0,2\p)$  \\
\hline
$k=n-1$ & $I_2^{\otimes n}+\proj{0}^{\otimes (n-1)}\otimes \diag(e^{i\phi}-1,e^{i\t}-1)$ & $\theta,\phi\in(0,2\pi)$ and $\theta\ne\phi$  \\
\hline
$k=2$ & $\proj{0}\otimes I_2^{\otimes (n-1)}+\proj{1}\otimes\diag(1,e^{i\b_2})\otimes\cdots\otimes\diag(1,e^{i\b_n})$ & $\b_2,\cdots,\b_n\in(0,2\pi)$  \\
\hline
$k=1$ & $\diag(\cos\a,1)\ox I_2^{\ox (n-1)}+i\sin\a\proj{0}\ox \sigma_3^{\ox (n-1)}$ & $\a\in(0,\frac{\pi}{2})\cup(\frac{\pi}{2},\pi)$  \\
\hline
$k=0$ & $\diag(\cos\a,\cos\b)\ox I_2^{\ox (n-1)}+i\diag(\sin\a,\sin\b)\ox\sigma_3^{\ox (n-1)}$ & $\a,\b\in(0,\frac{\pi}{2})\cup(\frac{\pi}{2},\pi)$ \\
\hline
\end{tabular}
\label{tab:nqubit}
\end{table*}

We analyze the entangling power of a $n$-qubit Schmidt-rank-two unitary acting on systems $A_1,A_2,...,A_n$. 
It suffices to consider the genuine multipartite unitaries where there are not product unitary operators across any bipartition. In fact, if the subsystems of a unitary are product unitaries then this part will not generate entanglement. When analyzing the entangling power of this unitary, we can always ignore these product unitaries  and turn to consider the non-product subsystems of this unitary. For example,  from Definition \ref{def:N-ep}, the entangling power of a unitary $U=(A_1\ox A_2+B_1\ox B_2)\ox A_3\ox...\ox A_n$ is equal to that of a smaller unitary $U_1=A_1\ox A_2+B_1\ox B_2$ as no entanglement is generated by the last $n-2$ subsystems.   

As is shown in TABLE \ref{tab:nqubit}, a genuine $n$-qubit ($n\geq 4$) unitary is classified into five categories based on its singular number. It is defined as  the  number of local singular operators in the unique Schmidt decomposition in \cite{shen2022class}. 
We obtain the entangling power of the $n$-partite unitary whose singular number $k$ is equal to $n,n-1,2,1,0$, respectively. 
\begin{proposition}
\label{pro:ep-nqubit}
Suppose $U_{k}$ is a Schmidt-rank-two $n$-qubit ($n\geq 4$) unitary with singular number $k$ in TABLE \ref{tab:nqubit}, for $k=n,n-1,2,1,0$. Then $U_k$ has the entangling power as follows.

(i) For $k=n$ or $n-1$, we assume that $\l_j$ is the $j$-th smallest elsment in the set $\big\{0,\t,\phi \big|\t\in [0,2\pi), \phi\in (0,2\pi)\big\}, \t\neq \phi$. Then
\begin{eqnarray}
&&K_E(U_k) \\
=&&
\begin{cases}
1, 
\;\mbox{for} \;\;
0\leq \l_j-\l_{j-1}\leq \pi,\;j=2,3
\;\mbox{and}\; \;
\l_{3}\geq \pi,
 \\
\max_{1\leq j \leq 3} H(\frac{1+\cos(\b)}{2},
\frac{1-\cos(\b)}{2}),\;\mbox{otherwise},
\end{cases}
\nonumber
\end{eqnarray}
where $\b=(\l_j-\l_{j+1\mod 3})/2$.

(ii) For $k=2$, 
we assume that $\t_j$ is the $j$-th smallest element in the set   $\big\{ \sum_{l=2}^n q_l\b_l\mod 2\pi  \big|q_l=0,1, \; \b_l\in (0,2\pi) \big\}$. Then
\begin{eqnarray}
\label{eq:ep k=2}
&& \!\!\!\! K_E(U_{k})\\
= && \!\!\!\!\begin{cases}
1, 
\;\mbox{for} \;
0\leq\t_j-\t_{j-1}\leq \pi,
\;
\t_{2^{n-1}}-\t_1\geq \pi, 
\\
\quad
j=2,3,...,2^{n-1}, \mbox{or}\;\exists t\in \{2,3,...,n\}\; \text{s.t.} \;\b_t=\pi, 
\\
\max_{\substack{1\leq j \leq 2^{n-1}\\
t\in \{2,3,...,n\}}}\Big\{  H(\frac{1+h_{jk}}{2},
\frac{1-h_{jk})}{2}),
H(\frac{1+g_t}{2},
\frac{1-g_t}{2})
\Big\}, \\
\hspace{6.2cm}
\mbox{otherwise},
\end{cases}
\nonumber
\end{eqnarray}
where $h_{jk}=\cos(\frac{\t_j-\t_k}{2})$ with  $k=j+1\mod 2^{n-1}$, and $g_t=\cos(\b_t/2)$.

(iii) For $k=1$ or $0$, we assume that $\l_j$ is the $j$-th smallest element in the set $\big\{(\pm2\a)\mod 2\pi, (\pm\b)\mod 2\pi\big|\a\in(0,\frac{\pi}{2})\cup(\frac{\pi}{2},\pi), \b\in[0,\frac{\pi}{2})\cup(\frac{\pi}{2},\pi) \big\}$.
 Then
\begin{eqnarray}
\label{eq:ep k=1,0}
&& \!\!\!\! K_E(U_{k})\\
= && \!\!\!\!
\begin{cases}
1, 
\;\mbox{for} \;
0\leq \l_j-\l_{j-1}\leq \pi,j=2,3,4,
\l_{4}-\l_1\geq \pi,
 \\
\max_{1\leq j \leq 4} H(\frac{1+\cos(\b)}{2},
\frac{1-\cos(\b)}{2}), 
\;\mbox{otherwise},
\end{cases}
\nonumber
\end{eqnarray}
where $\b=\frac{\l_j-\l_k}{2}$ with $k=j+1\mod 4$.
\end{proposition}
The proof of Proposition \ref{pro:ep-nqubit} is given in Appendix \ref{apx: n-qubit ep}.

\begin{figure}[h]
    \centering
    \includegraphics[width=9cm]
    {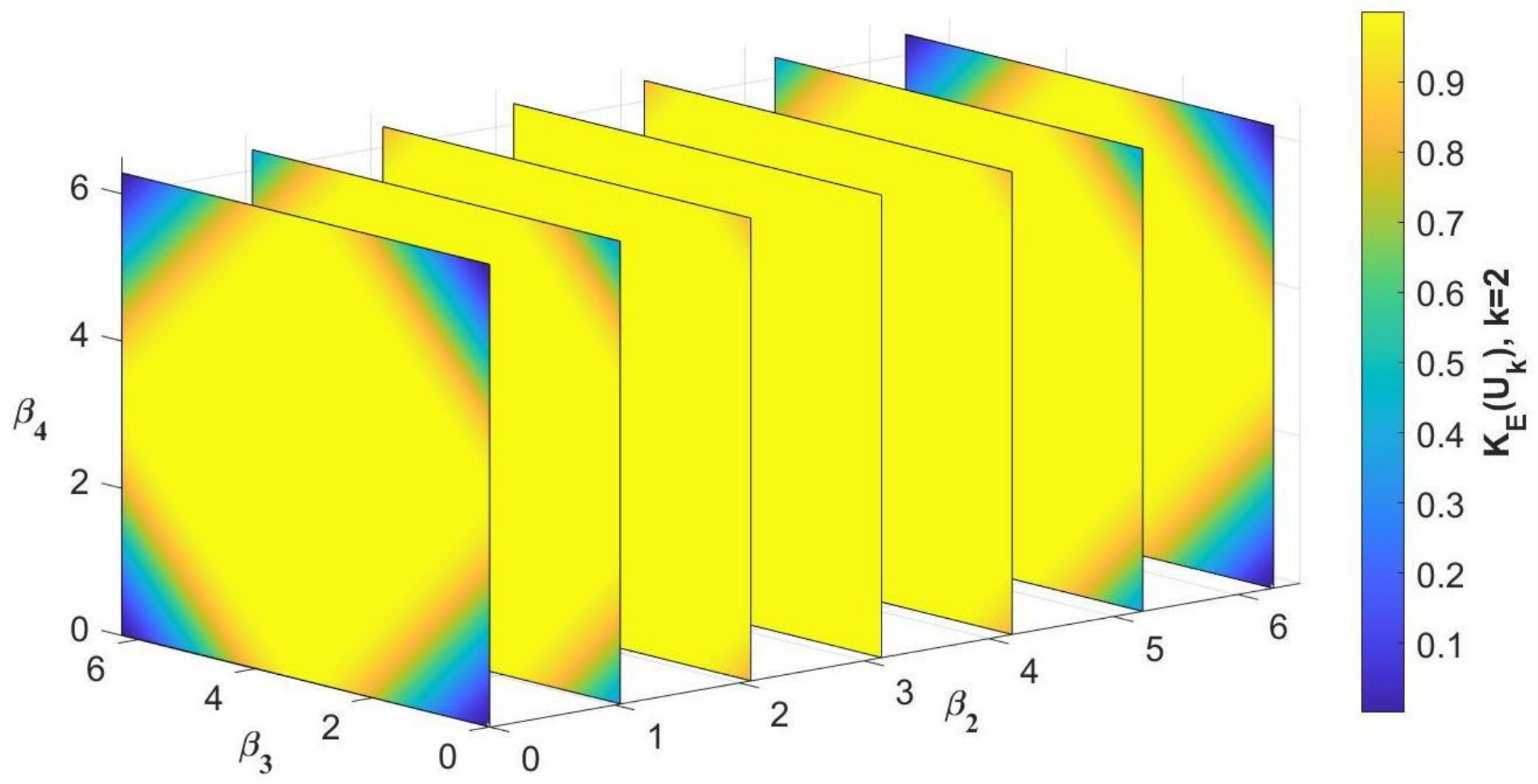}
    \caption{The entangling power of the Schmidt-rank-two four-qubit unitary with singular number $k=2$, that is, $U_2=\proj{0}\otimes I_2^{\otimes 3}+\proj{1}\otimes\diag(1,e^{i\b_2})\otimes\cdots\otimes\diag(1,e^{i\b_4})$. The entangling power of $U_2$ is shown with respect to the parameters $\b_2,\b_3,\b_4$, where $\b_2=\frac{k\pi}{3}$ with $k=0,1,...,6$ and $\b_3,\b_4\in(0,2\pi)$. Here the value of $K_E(U)$ is shown by the color corresponding to the colorbar.}
    \label{fig:Uk=2}
\end{figure}

We show the results of Proposition \ref{pro:ep-nqubit} in FIG. \ref{fig:Uk=n,1} and \ref{fig:Uk=2}. In detail, we show results (i) and (iii) in FIG. \ref{fig:Uk=n,1}, and result (ii) in FIG. \ref{fig:Uk=2}. From FIG. \ref{fig:Uk=n,1}, one can see that the value of $K_E(U_{n-1})$ reaches the maximum, i.e. one, when both $\phi$ and $\t$ approach $\pi$, and   $K_E(U_{n-1})$ tends to be the minimum, i.e. zero, when  $\phi$ or $\t$ approach $0$ or $2\pi$.  
When $\t\rightarrow 0$ (real line) or $\t\rightarrow \phi$ (dashed line),  $K_E(U_{n-1})$ approaches $K_E(U_{n})$, which is merely determined by $\phi$. 
Besides, one can obtain that  the value of $K_E(U_{0})$ reaches the maximum, i.e. one, when $\a(\text{resp. }\b)\in (\frac{\pi}{4},\frac{3\pi}{4})$, $\b(\text{resp. }\a)\in (0,\frac{\pi}{4})$ or $\a(\text{resp. }\b)\in (\frac{\pi}{4},\frac{3\pi}{4})$, $\b(\text{resp. }\a)\in (\frac{3\pi}{4},\pi)$. Besides, $K_E(U_{n-1})$ approaches the minimum, zero, when $\phi$ or $\t$ approach 0 or $\pi$, or  both of them approach $\pi$. 
 When $\b\rightarrow 0$ (real line) or $\b\rightarrow \pi$ (dashed line), the entangling power of $U_{0}$ approaches that of $U_{1}$. 
 
 From FIG. \ref{fig:Uk=2}, the value of $K_E(U_2)$ is shown regarding to the parameters $\b_2,\b_3,\b_4$. From the proof of Proposition \ref{pro:ep-nqubit}, $\b_2,\b_3,\b_4$ have the same effect on $K_E(U_2)$. Without loss of generality, we set $\b_2=\frac{k\pi}{3}$ with $k=0,1,...,6$ such that we can observe  the value of $K_E(U_2)$ clearly. First we show the value of $K_E(U_2)$  regarding $\b_2$.
 From the first plane on the left, one can see that  the minimum of $K_E(U_2)$ approaches zero for $\b_2\rightarrow 0$. The minimum of $K_E(U_2)$ increases until $b_2$ reaches $\pi$, where $K_E(U_2)=1$ for any $\b_3,\b_4$, shown in the third plane. After that, the minimum of $K_E(U_2)$ decreases with the increase of $\b_2$, shown by the fifth to seventh planes. Besides, the maximum of $K_E(U_2)$ remains one for any $\b_2\in (0,2\pi)$. Second,  we show the value of $K_E(U_2)$ for a fixed $\b_2$ and variable $\b_3,\b_4$. The value of $K_E(U_{2})$ reaches the maximum, i.e. one, when both $\b_3$ and $\b_4$ approach $\pi$, and   $K_E(U_{2})$ tends to be the minimum, i.e. zero, when  $\b_3$ or $\b_4$ approach $0$ or $2\pi$.

\begin{figure}[htbp]
\centering
\subfigure{
\includegraphics[width=6cm]{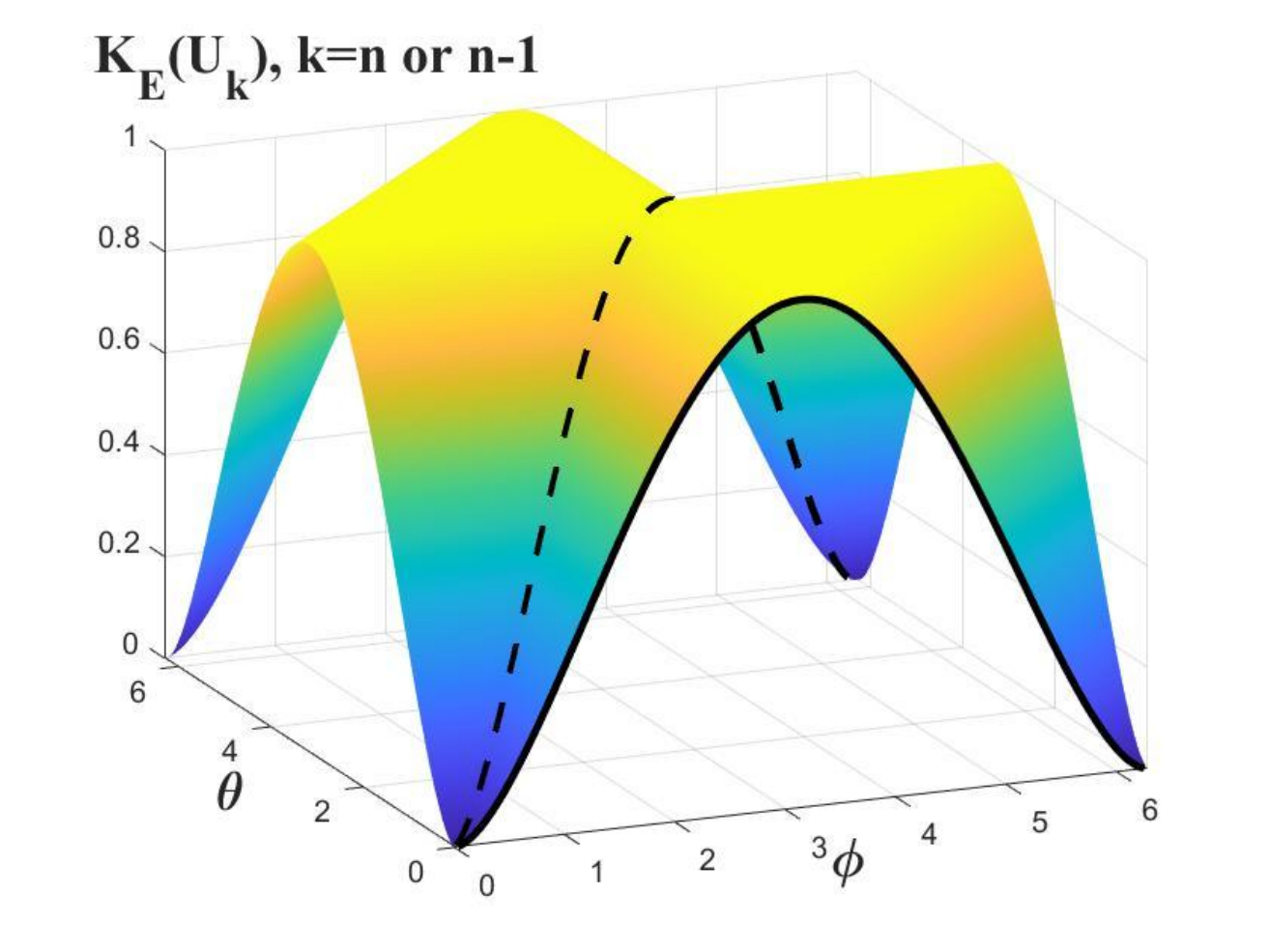}
}
\subfigure{
\includegraphics[width=6cm]{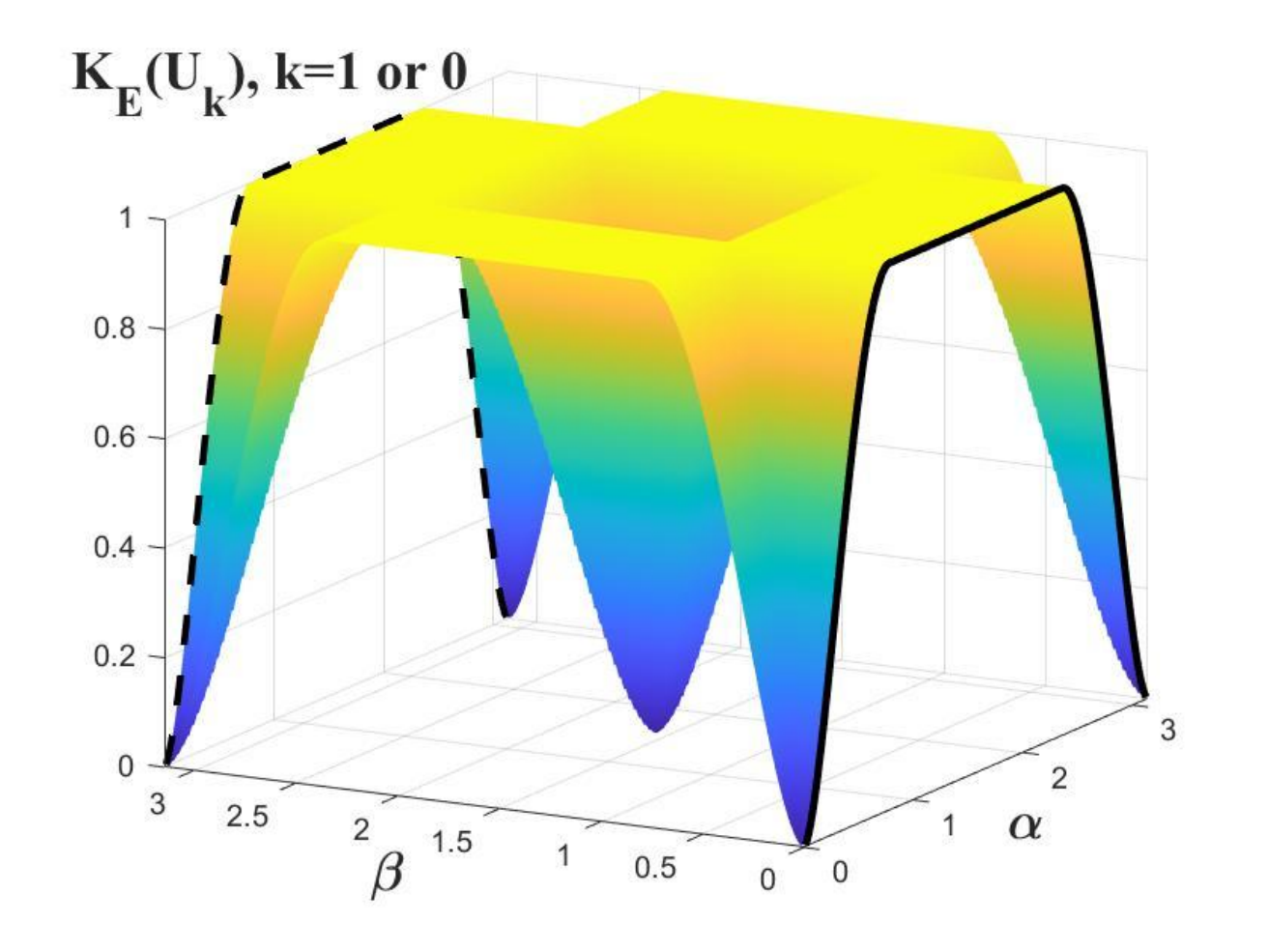}
}
\caption{ The entangling power of the Schmidt-rank-two $n$-qubit unitary with singular number $n,n-1,1,0$, i.e. $U_{n}$, $U_{n-1}$ (LHS), and $U_{1}$,  $U_{0}$ (RHS).
(a) The entangling power of $U_{n}$ in (\ref{def:Un}) is shown by the real and dashed line, with regard to  the parameter $\phi$.    The entangling power of $U_{n-1}$ in (\ref{def:Un-1}) is shown with respect to the parameters $\phi$ and $\t$. 
(b) The entangling power of $U_{1}$ in (\ref{def:Un}) is shown by the real and dashed line, with regard to  the parameter $\a$. The entangling power of $U_{0}$ in (\ref{def:U0}) is shown with respect to the parameters $\a$ and $\b$.}
\label{fig:Uk=n,1}
\end{figure}

We show  the assisted entangling power of some Schmidt-rank-two multi-qubit unitaries by Propositions \ref{pro: aep max} and \ref{pro:ep-nqubit}. Here  the assisted entangling power reaches the maximum.
\begin{proposition}
\label{pro:aep-nqubit}
Suppose $U_{k}$ is a Schmidt-rank-two $n$-qubit ($n\geq 4$) unitary with singular number $k$ in TABLE \ref{tab:nqubit}, for $k=n,n-1,2,1,0$. Then   the assisted entangling power of   $U_k$ reaches the maximum, i.e. one ebit if and only if 

(i) For $k=n$ or $n-1$, 
we assume that $\l_j$ is the $j$-th smallest element in the set $\big\{0,\t,\phi \big|\t\in [0,2\pi), \phi\in (0,2\pi)\big\}, \t\neq \phi$.
It holds that $0\leq \l_j-\l_{j-1}\leq \pi$, $j=2,3$
and $\l_{3}\geq \pi$.

(ii) For $k=2$, 
we assume that $\t_j$ is the $j$-th smallest element in the set   $\big\{ \sum_{l=2}^n q_l\b_l\mod 2\pi  \big|q_l=0,1, \; \b_l\in (0,2\pi) \big\}$.
It holds that 
$0\leq\t_j-\t_{j-1}\leq \pi$
 and 
$\t_{2^{n-1}}-\t_1\geq \pi$,
for $j=2,3,...,2^{n-1}$,
or there is a $t\in \{2,3,...,n\}$ such that $\b_t=\pi$.

(iii) For $k=1$ or $0$, we assume that $\l_j$ is the $j$-th smallest element in the set $\big\{(\pm2\a)\mod 2\pi, (\pm\b)\mod 2\pi\big|\a\in(0,\frac{\pi}{2})\cup(\frac{\pi}{2},\pi), \b\in[0,\frac{\pi}{2})\cup(\frac{\pi}{2},\pi) \big\}$.
It holds that
$0\leq \l_j-\l_{j-1}\leq \pi$, for $j=2,3,4$ and 
$\l_{4}-\l_1\geq \pi$.
\end{proposition}
\begin{proof}
The proof of Proposition \ref{pro:ep-nqubit} shows that the unitaries in case (i)-(iii) are in the form of two-term control by any bipartition. Then the result can be obtained in a similar way as the proof of Proposition \ref{pro:ep3qubit}.
\end{proof}

Proposition \ref{pro:aep-nqubit} shows the necessary and sufficient condition that the Schmidt-rank-two multi-qubit assisted entangling power reaches the maximum. Many well-known multi-qubit universal quantum gates are included in  Proposition \ref{pro:aep-3qubit}, such as the $n$-qubit Toffoli gate, $n$-qubit controlled-controlled z gate, and the generalized CNOT gate proposed in \cite{lcly20160808} etc. In general, the multipartite assisted entangling of an arbitrary unitary operation is still an open problem.

\section{Entangling and assisted entangling power of Widely-used multi-qubit unitaries}
\label{sec:ep widely used}
In this section we consider the entangling power of two types of multi-qubit unitary gates, namely Toffoli gates and Fredkin gates. We show that the entangling and assisted entangling power of a $n$-qubit Fredkin gate is equal to one ebit, regardless of the number of controlling parties. As for the Fredkin gate, we show that the entangling power of three-qubit Fredkin gate is equal to two ebits, and conjecture that the entangling power of four-qubit Fredkin gate may also be equal to two ebits.

The Toffoli gate or controlled-controlled-NOT gate is a three-qubit gate. Its circuit representation is shown in FIG. \ref{fig:TF gates}. Toffoli gate is a kind of universal reversible gate. This gate acts as follows: the two control bits are unchanged, i.e. $a'=a$ and $b'=b$ while the target bit is flipped if and only if  the two control bits are set to 1, i.e. $c'=c\oplus ab$.
The expression of the three-qubit Toffoli gate is given by 
\begin{eqnarray}
\label{def:toffoli}
T_3=(I_2^{\ox 2}-\proj{1,1})\ox I_2+\proj{1,1}\ox \s_x.
\end{eqnarray}
Note that the controlled-controlled z (CCZ) gate, a special case for the controlled-controlled phase (CCP) gate, 
\begin{eqnarray}
CCP=(I_2^{\ox 2}-\proj{1,1})\ox I_2+\proj{1,1}\ox \diag(1,e^{i\phi})
\nonumber
\end{eqnarray}
for  $\phi\in (0,2\pi)$
is equivalent to the Toffoli gate up to local unitaries. Their entangling power is obtained as they are Schmidt-rank-two unitaries with singular number equal to $n$. From (i) in Proposition \ref{pro:ep-nqubit}, the entangling power of a CCP gate is equal to $H(\frac{1+\cos(\phi/2)}{2}, \frac{1-\cos(\phi/2)}{2})$ with $\phi\in (0,2\pi)$. In particular, by choosing $\phi=\pi$, the CCZ and Toffoli gate is equal to one ebit, which is the maximal of $CCP$ gate. 

In general, we may consider the entangling power of a $n$-qubit Toffoli gate $T_n$, namely
\begin{eqnarray}
\label{def:toffolin}
T_n=(I_2^{\ox (n-1)}-\proj{1}^{\ox n})\ox I_2+\proj{1}^{\ox n} \ox \s_x.
\end{eqnarray}
By Proposition \ref{pro:ep-nqubit}, it can be obtained that the entangling power of $T_n$ is also one ebit, which will not increase with the number of controlling parties. This derives the following fact.
\begin{proposition}
\label{pro: n-toffoli ep}
The entangling and assisted entangling power of a $n$-qubit Toffoli gate $T_n$ is equal to one ebit.    
\end{proposition}
	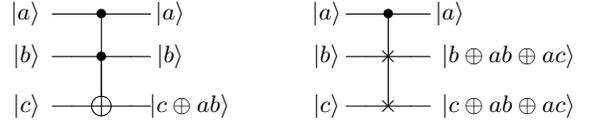
\begin{figure}[h!] 
	\centerline{
		\Qcircuit @C=0.8em @R=1.5em {
\lstick{\ket{a}}  &\qw&\ctrl{2}&\qw &\qw&\mbox{$\ket{a}$}& & & & & & & &\mbox{$\ket{a}$}& &\qw &\ctrl{1} &\qw&\qw& \mbox{$\ket{a}$}\\
\lstick{\ket{b}}  &\qw&\ctrl{1}&\qw&\qw&\mbox{$\ket{b}$}& & & & & & & & \mbox{$\ket{b}$}& &\qw &\qswap   &\qw&\qw& & & &\mbox{$\ket{b\oplus ab\oplus ac}$}\\
\lstick{\ket{c}}  &\qw&\targ &\qw&\qw &  &\mbox{$\ket{c\oplus ab}$} & & & & & & &\mbox{$\ket{c}$} & &\qw &\qswap\qwx &\qw&\qw & & & &\mbox{$\ket{c\oplus ab\oplus ac}$} \\
		}
	}
	\caption{The circuit representation for the three-qubit Toffoli gate (LHS) and Fredkin gate (RHS). The qubits on LHS are inputs of the circuit and those on RHS are outputs, where $a,b,c\in \{0,1\}$. }
	\label{fig:TF gates}
\end{figure}

The Fredkin gate, or controlled-swap gate, is another universal reversible gate, whose circuit representation is shown in FIG. \ref{fig:TF gates}. This gate swaps the input bits $b$ and $c$ if and only if the control bit $a$ is set to 1. The expression of the three-qubit Fredkin gate is given by
\begin{eqnarray}
F_3=\proj{0}\ox I_2^{\ox 2}+\proj{1}\ox S_2,
\end{eqnarray}
where $S_2$ is the two-qubit swap gate. 
Next we consider the entangling power of the Fredkin gate $F_3$ acting on $\cH_A\ox \cH_B\ox\cH_C$, and obtain the following.
\begin{proposition}
\label{pro: 3-fredkin ep}
The entangling power of a three qubit Fredkin gate $F_3$ is equal to two ebits.    
\end{proposition}
\begin{proof}
 Since system $B$ is symmetric with system $C$ here, it suffices to consider the bipartition $A:BC$ and $C:AB$.  

First we consider the entanglement generation $K_{A:BC}(F_3)$. Obviously $F_3$ can be controlled from system $A$. From Proposition \ref{pro: ep max} and Lemma \ref{le:triepmax}, the entanglement generation $K_{A:BC}(F_3)$ can be derived by removing the ancilla $R_A$, that is,
\begin{eqnarray}
&& \!\!\!\!\!\! K_{A:BC}(F_3)\\
=&& \!\!\!\!\!\!
\max_{\substack{\ket{\b},\ket{\g}\in \cH_{BR_B}\ox\cH_{CR_C}\\p\in [0,1]}  }S(p\proj{\b}+(1-p)S_2\proj{\g}S_2).
\nonumber
\end{eqnarray}
Since $F_3$ is controlled by two terms from system $A$, we have $K_{A:BC}(U)\leq \log_2 2=1$ (ebit).
By choosing $\ket{\b}=\ket{00}_{BR_B}\ket{00}_{CR_C}$ and $\ket{\g}=\ket{00}_{BR_B}\ket{10}_{CR_C}$, one has $K_{A:BC}(U)\geq \max_{p\in [0,1]} S(p \proj{00}_{BC}+(1-p)\proj{10}_{BC})=1$ (ebit). Hence   
$K_{A:BC}(F_3)=1 \text{ ebit}$.  

Next we consider the entanglement generation $K_{AB:C}(F_3)$.
  From the Schmidt decomposition of Fredkin gate under the bipartition $AB:C$, one can see that $K_{C:AB}(F_3)\leq \log_2 \sch(F_3)=2$. Next we choose $\ket{\a}_{AR_A}=\ket{1,0}$ and $\ket{\b}_{BR_B}=\ket{\g}_{CR_C}=\frac{\ket{00}+\ket{11}}{\sqrt{2}}$. Then 
$K_{C:AB}(F_3)\geq  E_{C:AB}(F_3(\ket{\a}\ox \ket{\b}\ox \ket{\g}))
=\log_2 4=2 \text{ (ebits)}$.
We have 
 $K_{C:AB}(F_3)=K_{B:AC}(F_3)=2 \text{ (ebits)}$.    
Hence one can obtain that 
\begin{eqnarray}
K(F_3)=&& \!\!\!\!\!\!
\max\{  K_{A:BC}(F_3),  K_{B:AC}(F_3),  K_{C:AB}(F_3)\}
\nonumber\\=&& \!\!\!\!\!\!
2 \text{ (ebits)}.\nonumber
\end{eqnarray}
This completes the proof.
\end{proof}

Next we consider the entangling power of $F_4$, where for the SWAP gate $S_2$,
\begin{eqnarray}
F_4=&& \!\!\!\!\!\!(\proj{00}+\proj{01}+\proj{10})_{AB}\ox
(I_2^{\ox 2})_{CD}
\nonumber\\
+&& \!\!\!\!\!\!
\proj{11}_{AB} \ox (S_2)_{CD}.
\nonumber
\end{eqnarray}
 Due to the symmetry of subsystems,  it suffices to consider four cases with respect to the bipartition.

Case 1: We analyze the bipartition $A:BCD$ or $B:ACD$, where $K_{A:BCD}(F_4)=K_{B:ACD}(F_4)$. We merely consider the bipartition $A:BCD$. By this bipartition, 
one can see that $\sch(F_4)=2$ and hence $K_{A:BCD}(F_4)=K_{B:ACD}(F_4)\leq 1$.
On the other hand, we have
\begin{eqnarray}
K_{A:BCD}(F_4)=\max_{p\in [0,1],\;\ket{\b},\ket{\g}\in \cH_{BR_B}\ox\cH_{CR_C}  \ox \cH_{DR_D} }
\nonumber\\
\times S(p\proj{\b}+(1-p)(I_2\ox S_2)\proj{\g}(I_2\ox S_2)).
\nonumber
\end{eqnarray}
We choose $\ket{\b}=\ket{00}_{BR_B}\ket{00}_{CR_C}\ket{00}_{DR_D}$ and $\ket{\g}=\ket{10}_{BR_B}\ket{00}_{CR_C}\ket{10}_{DR_D}$. Then $K_{A:BCD}(F_4)\geq 1$. To conclude, we have $K_{A:BCD}(F_4)=K_{B:ACD}(F_4)= 1$ (ebit). 

Case 2: We analyze the bipartition $C:ABD$ or $D:ABC$, where $\sch(F_4)=4$. It can be obtained that $K_{C:ABD}(F_4)=K_{D:ABC}(F_4)\leq 2$. Next we consider the bipartition $D:ABC$. By choosing the input state as $\ket{\a}_{AR_A}=\ket{\b}_{BR_B}=\ket{10}$, $\ket{\g}_{CR_C}=\ket{\eta}_{DR_D}=\frac{\ket{00}+\ket{11}}{\sqrt{2}}$. One can obtain that 
\begin{eqnarray}
K_{D:ABC}(F_4)\geq&& \!\!\!\!\!\!
E_{D:ABC}(F_4(\ket{\a}\ox \ket{\b}\ox \ket{\g}\ox \ket{\eta}))\nonumber\\
=&& \!\!\!\!\!\!
\log_2 4=2 \text{ ebits}. 
\nonumber
\end{eqnarray}
Therefore, we obtain that $K_{C:ABD}(F_4)=K_{D:ABC}(F_4)= 2$ (ebits).

Case 3: We analyze the bipartition $AB:CD$. By this bipartition, $F_4$ is a controlled unitary and it is controlled with two terms. Here systems $AB$ are the controller. From Lemma \ref{le:triepmax}, one has 
\begin{eqnarray}
K_{AB:CD}(F_4)=&& \!\!\!\!\!\!\max_{p\in [0,1],\;\ket{\b},\ket{\g}\in \cH_{CR_C}\ox\cH_{DR_D}  }\nonumber\\
 && \times S(p\proj{\b}+(1-p)S_2\proj{\g}S_2). \nonumber
\end{eqnarray}
By choosing the input state $\ket{\b}=\ket{00}_{BR_B}\ket{00}_{CR_C}$ and $\ket{\g}=\ket{00}_{BR_B}\ket{10}_{CR_C}$, one has $K_{A:BC}(U)\geq \max_{p\in [0,1]} S(p \proj{00}_{BC}+(1-p)\proj{10}_{BC})=1$ (ebit). Obviously it reached the maximum.  Hence we have $K_{AB:CD}=1$ (ebit).  

 We shall emphasize that the analysis in Case 1-3 can be extended to a $n$-qubit Fredkin gate. Next we consider the last bipartition.

Case 4: We consider the entanglement generation $K_{AD:BC}(F_4)$ or equivalently, $K_{AC:BD}(F_4)$. The Schmidt decomposition of $F_4$ with respect to the bipartition $AD:BC$ shows that $\sch(F_4)=5$ and thus $K_{AD:BC}(F_4)\leq \log_2 5$ (ebits).
On the other hand, we can choose an input state $\ket{\eta}=\ket{10}_{AR_A} \ox \ket{10}_{BR_B} \ox (\frac{\ket{00}+\ket{11}}{\sqrt{2}})_{CR_C} \ox (\frac{\ket{00}+\ket{11}}{\sqrt{2}})_{DR_D}$ such that 
$E_{AD:BC}(F_4\ket{\eta})=2$ (ebits).  Hence we have $K_{AD:BC}(F_4)\geq 2$, namely 
$K_{AD:BC}(F_4)\in [2,\log_25]$.
Further we conjecture that $K_{AD:BC}(F_4)\leq 2$ by numerical results and thus have the following.
\begin{conjecture}
\label{coj:KAD:BC}
The entanglement generation $K_{AD:BC}(F_4)=2$. Hence the entangling power of a four-qubit Fredkin gate $F_4$ is equal to two ebits. 
\end{conjecture}
In order to prove this conjecture, we may choose the input states as 
\begin{eqnarray}
	\label{eq:a1a2a3a4}
	&&\ket{\a_1}_{AR_A}=[w_0, w_1, w_2, 0]^T, \;
	\ket{\a_2}_{BR_B}=[ y_0, y_1,y_2, 0]^T, \;
 \nonumber\\
	&&\ket{\a_3}_{CR_C}=[ z_0, z_1, z_2, 0]^T, \;
	\ket{\a_4}_{DR_D}=[ 
 0, x_1, x_2, 0]^T.
 \nonumber
\end{eqnarray}
up to local unitaries. The subadditivity of von Neumann entropy may also be used. The entangling power of a $n$-qubit Fredkin gate $T_n$ remains to be an open problem.

\section{Conclusion}
\label{sec:conclusion}
In summary, we have introduced the definition of the entangling power of multipartite nonlocal unitary operations.  We have shown that the entangling, assisted entangling and disentangling power are assumed by an input state by some facts in functional analysis. 
The entangling power of Schmidt-rank-two multi-qubit unitary operations has been analytically derived. 
The necessary and sufficient condition that the assisted entangling power of Schmidt-rank-two multi-qubit unitary operations reaches the maximum has been given.
Further the entangling and assisted entangling power of the most widely-used multi-qubit unitary gates including Toffoli and Fredkin gates have also been investigated.

Many problems arising from this paper can be further explored. 
The entangling power of multipartite unitary operation that is controlled with more than two terms may be derived by the investigation between von Neumann entropy and the reduced density matrix of the controlled systems. 
Besides, the entangling power of multipartite unitary operations with Schmidt rank more than two remains to be analyzed. This may be obtained by the canonical form of such unitary operations. Some interesting problems concerning this problem remain to be analyzed, such as Conjecture \ref{coj:KAD:BC}.
As a harder problem, the analytical results of the multipartite assisted entangling power of an arbitrary unitary operation  remains to be investigated.

\section*{ACKNOWLEDGMENTS}
Authors were supported by the NNSF of China (Grant No. 12471427).

\appendix
\section{Proof of properties of entangling, assisted entangling and disentangling power}
\label{apx: property}

Here we show the proof of Lemma \ref{le: p-norm compact}.

\begin{proof}
(i) Let $\cS$ denote the set consisting of all density operators of $(d_1d_2...d_n)$-dimensional $n$-partite pure states. So $\cS$ is a subset of the normed space $(\cX_n,\norm{\cdot}_p)$.
From (1) in Theorem \ref{th:fdns compact}, it suffices to prove that $\cS$ is bounded and closed. Obviously $\cS$ is bounded, as $\r=\proj{\ps}\in \cS$ satisfies $\norm{\r}_p=1$ for $p\in[1,+\infty]$. 
In order to show that $\cS$ is closed under the distance $d_p(\cdot,\cdot)$ induced by $\norm{\cdot}_p$, we need to show that all the accumulation points of $\cS$ belong to $\cS$. 
We consider a sequence of $n$-partite pure states $\{\r_k=\proj{\ps_k}\}_{k=1}^\infty$, where $\ket{\ps_k}=[a_j^{(k)}]_{j=1}^{d_1...d_n}$ satisfies that $\lim_{k\rightarrow\infty}a_j^{(k)}=a_j$ exists for all $j$. That is, $\lim_{k\rightarrow\infty} d_p(\r_k,\r)=0$, where $\r=\textbf{a}\cdot\textbf{a}^\dg$ with $\textbf{a}=[a_j]_{j=1}^{d_1...d_n}$.  From the identity $\tr[\r_k]=\tr[\r_k^2]=1$ we obtain that $\lim_{k\rightarrow\infty}\tr[\r_k]=\tr[\r]=1$ and  $\lim_{k\rightarrow\infty}\tr[\r_k^2]=\tr[\r^2]=1$. It implies that if $\{\r_k\}_{k=1}^\infty\subset \cS$ and $\lim_{k\rightarrow\infty} d_p(\r_k,\r)=0$ then $\r\in \cS$. So $\cS$ is closed under $d_p(\cdot,\cdot)$.

(ii)  Suppose $\cP$ is a subset of this normed space consisting of 
		all density operators of  $d_1d_2...d_n$-dimensional $n$-partite product states. 
		In order to obtain that $\cP$ is compact, it suffices to show its boundedness and closeness.  Any $\r'\in \cP$ satisfies $\norm{\r'}_p=1$ and thus $\cP$ is bounded.
		We show that $\cP$ is closed under $d_p(\cdot,\cdot)$. 
		Consider a sequence of bipartite product states $\{\r'_k=\proj{\ps'_k}\}_{k=1}^\infty$, with $\ket{\ps'_k}=\otimes_{j=1}^n\ket{\a_j^{(k)}}= \otimes_{j=1}^n[a_{j,m_j}^{(k)}]_{m_j=1}^{d_j}$ where $\lim_{k\rightarrow\infty}a_{j,m_j}^{(k)}=a_{j,m_j}$ exist for all $j,m_j$. That is, $\lim_{k\rightarrow\infty} d_p(\r'_k,\r')=0$, where $\r'=(\textbf{a}_1\otimes\textbf{a}_2\otimes...\otimes\textbf{a}_n)(\textbf{a}_1\otimes\textbf{a}_2\otimes...\otimes\textbf{a}_n)^\dg:=\r'_{A_1}\otimes\r'_{A_2}...\otimes\r'_{A_n}$ with $\textbf{a}_j=[a_{j,m_j}]_{m_j=1}^{d_j}$ and $j=1,2,...,n$. From the identity $\tr_{X_t}[\r'_k]=\tr_{X_t}[(\r'_k)^2]=1$ with $X_t=\{A_1A_2 ... A_n\} / \{A_t\}$ and $t=1,2,...,n$, we obtain that $\lim_{k\rightarrow\infty}\tr_{X_t}[\r'_k]=\tr[\r'_{A_t}]=1$ and  $\lim_{k\rightarrow\infty}\tr_{X_t}[(\r'_k)^2]=\tr[(\r'_{A_t})^2]=1$.  Hence, $\r'\in \cP$, which  implies that $\cP$ is closed under $d_p(\cdot,\cdot)$ for $p\in[1,+\infty]$. 
		Using Theorem \ref{th:fdns compact}, $\cP$ is a compact subset of the normed space $(\cX_2,\norm{\cdot}_p)$. 
\end{proof}

The proof of Proposition \ref{pro: aep max} is shown below.

\begin{proof}
(i) By the definition of $K_{E_a}(U)$, it is derived by taking the maximum over 
all possible bipartition and the supreme of all pure states.
As the bipartite cuts has finite kinds, the maximum can be reached by one of the bipartition, namely $A:B$. It suffices to show that the supreme can be reached by this bipartition  $A:B$. Without loss of generality, we treat this multipartite unitary $U$ as a bipartite one, i.e. $U_{AB}$.

We consider the normed space $(\cX_2,\norm{\cdot}_1)$, where  $\cX_2$ is the set of bipartite operators given by $M=\sum_{i,j=1}^{d_Ad_{R_A}} \ket{i}\bra{j}\otimes M_{i,j}$ with $M_{i,j}\in \cM_{d_Bd_{R_B}}$. 
Lemma \ref{le: p-norm compact} shows that  all density operators of  $d_Ad_{R_A}\times d_Bd_{R_B}$-dimensional bipartite pure states form a compact subset $\cS$ of this normed space.
Next we consider the following mapping 
\begin{eqnarray}
	\label{def: aepmap}
	K_{E_a}^U: \cS\rightarrow \bbR,
 \quad \r\mapsto S(\tr_{AR_A} U\r U^\dg)-S(\tr_{AR_A}  \r ),
\end{eqnarray}
where $\r\in \cS$, $U\in \cU_{d_Ad_B}$ is a bipartite unitary. From Theorem \ref{cor:cm}, if  $K_{E_a}^U$ is a continuous mapping of $\cS$  then the assertion is proved.
It suffices to show that $K_{E_a}^U$ is continuous at every point $\r_0\in\cS$. For every $\ve>0$, we show the existence of a $\d>0$ such that $\abs{ K_{E_a}^U(\r)-K_{E_a}^U(\r_0)}<\ve$ for all $\r\in\cS$ satisfying $d_1(\r,\r_0)<\d$.                                                                                                                                                   
In detail, for above $\ve>0$,  there are $\d_1=\min\{\frac{\ve}{4\log (d_Ad_{R_A}d_Bd_{R_B})},\frac{1}{e} \}>0$ and some points $\r\in\cS$ satisfying $d_1(\r,\r_0)=d_1(U\r U^\dg,U\r_0U^\dg)<\d_1$, where the equality follows from the unitary invariance of trace norm. Since the trace norm is non-increasing under the action of partial trace, it holds that $\max\big\{d_1(\tr_{AR_A}\r ,\tr_{AR_A}\r_0),d_1(\tr_{AR_A}U\r U^\dg,\tr_{AR_A}U\r_0U^\dg)\big\}<\d_1$. Using triangle inequality and Fannes' inequality of von Neumann entropy \cite{nielsen2000quantum}, one can obtain that 
\begin{eqnarray}
&&\!\!\!\!\!\!\abs{ K_{E_a}^U(\r)-K_{E_a}^U(\r_0)}
\nonumber\\
\leq &&\!\!\!\!\!\!
\abs{S(\tr_{AR_A} U\r U^\dg)-S(\tr_{AR_A} U\r_0 U^\dg)}
\nonumber\\
&&\!\!\!\!\!\!
+\abs{S(\tr_{AR_A}\r)-S(\tr_{AR_A}\r_0)}
\nonumber\\
\leq&& \!\!\!\!\!\! 2(\d_1\log (d_Ad_{R_A}d_Bd_{R_B})+\eta(\d_1)),
\nonumber
\end{eqnarray}
where $\eta(x):=-x\log x$. Since  $\lim_{x\rightarrow 0}\eta(x)=0$, for above $\ve>0$, there is a $\d_2>0$ such that $0<\eta(x)<\ve/4$ for all $x$ satisfying $0<x<\d_2$. We can always find a  $\d=\min\{\d_1,\d_2\}$ such that 
\begin{eqnarray}
	\label{eq:leq2delta}
&&\!\!\!\!\!\!\abs{ K_{E_a}^U(\r)-K_{E_a}^U(\r_0)}
\\
<&&\!\!\!\!\!\! 
2(\d\log (d_Ad_{R_A}d_Bd_{R_B})+\eta(\d))<\ve
\end{eqnarray}
for all $\r\in\cS$ satisfying $d_1(\r,\r_0)<\d$. Here the first inequality follows from $\d<1/e$ and Fannes' inequality. The second inequality comes from $\d=\min\{\d_1,\d_2\}$, i.e.  $\d\leq\d_1\leq\frac{\ve}{4\log (d_Ad_{R_A}d_Bd_{R_B})}$  and $\d\leq \d_2$.
 Hence, the mapping $K_{E_a}^U$ is continuous by Definition \ref{def: continuous map}.
From Theorem \ref{cor:cm}, the continuous mapping $K_{E_a}^U$ assumes a maximum at some points of compact subset $\cS$.

(ii) The assertion is obtained by replacing the unitary $U$ in (i) by $U^\dg$. This completes the proof.
\end{proof}

Below we show the proof of Proposition \ref{pro: ep max}.

\begin{proof}
As the proof of Proposition \ref{pro: aep max}, it suffices to consider the multipartite unitary under the bipartition $A:B$.  
In the normed space $(\cX_2,\norm{\cdot}_1)$ defined above, we consider a subset $\cP$ consisting of 
 all density operators of  $d_Ad_{R_A}\times d_Bd_{R_B}$-dimensional bipartite pure product states. In Lemma \ref{le: p-norm compact}, we have shown that $\cP$ is a compact subset of this normed space.  On the other hand,  similar to the proof of Proposition \ref{pro: aep max}, we can prove that  $K_{E}^U$ is a continuous mapping of $\cP$  for any bipartite unitary $U\in \cU_{d_Ad_B}$, where for $\r'\in \cP$,
\begin{eqnarray}
	&&K_{E}^U: \cP\rightarrow \bbR,
 \quad
 \r'\mapsto S(\tr_{AR_A} U\r' U^\dg).
\end{eqnarray}
From Theorem \ref{cor:cm}, the continuous mapping $K_{E}^U$ assumes a maximum at some points of compact subset $\cP$.
\end{proof}
 
\section{Proof of preliminary lemmas of Schmidt-rank-two bipartite unitary entangling power}
\label{apx:pre lemma}
The proof of Lemma \ref{le: miniconv} is shown here.

\begin{proof}
Geometrically, $e^{i\t_j}$ stands for a point on the unit circle in the complex plane. The minimization can be transformed to the minimum distance between the original point and convex hull of $\cC$, denoted by $\mbox{conv}(\cC)$.
In the complex plane, $\mbox{conv}(\cC)$ is expressed as a $n$-polygon and its interior, whose vertices belong to $\cC=\{e^{i\t_j}\}_{j=1}^{n}$. 
	If the original point $O\in \mbox{conv}(\cC)$, the minimum distance is equal to zero; otherwise the distance is equal to the minimum distance between $O$ and each side of $\mbox{conv}(\cC)$, i.e. $\min_{1\leq j\leq n} \abs{\cos(\frac{\t_j-\t_{j+1\mod n}}{2})}$. The two cases are shown in FIG \ref{fig:schu=21}.
 
	\begin{figure}[h]
\center{\includegraphics[width=8cm]  {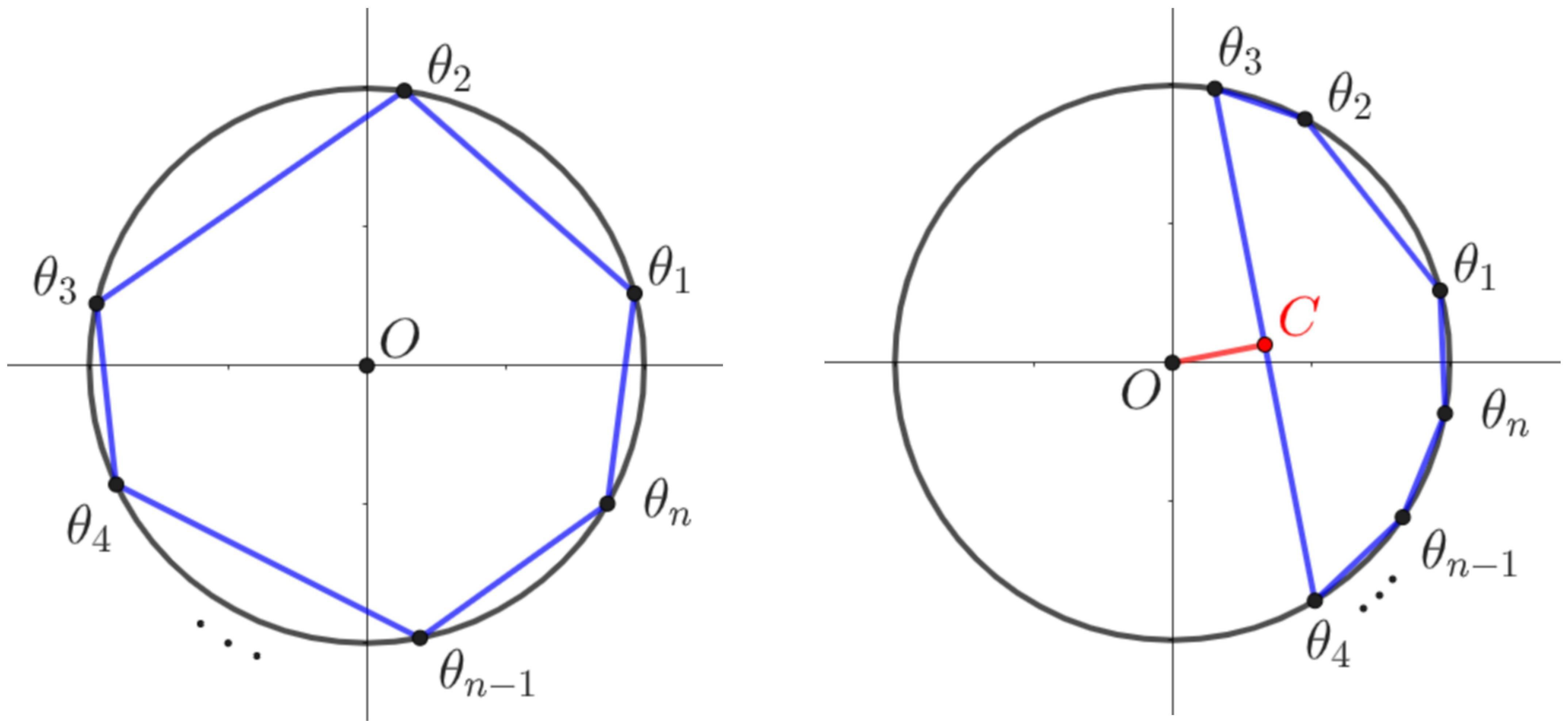}}
		\caption{The polygon and its interior represents $\mbox{conv}(\cC)$. If $O\in \mbox{conv}(\cC)$ then the minimization is equal to zero (LHS), otherwise it is equal to  $\abs{OC}=\abs{\cos(\frac{\t_4-\t_3}{2})}$ (RHS).  }
		\label{fig:schu=21}
	\end{figure}
	So it suffices to consider when the original point belongs to  $\mbox{conv}(\cC)$. In fact, $n$ boundary situations should be considered, where one of the edges of this $d_B$-polygon becomes the diameter of unit circle.
	In the boundary situations, the point $O$ is on one of the edges of this polygon. One can obtain that the boundary situations are depicted by $\t_j-\t_{j-1}=\pi$ with $j=2,3,...,n$ and $\t_{n}-\t_1=\pi$. They are shown in FIG. \ref{fig:schu=22}.   One can see that $O\in\mbox{conv}(\cC)$ if and only if all edges of the polygon are not on the same half of unit circle. That is $\t_j-\t_{j-1}\leq\pi$ and $\t_{n}-\t_1 \geq\pi$, for $j=2,3,...,n$. This completes the proof. 
	\begin{figure*}[htp]
\center{\includegraphics[width=13cm]  {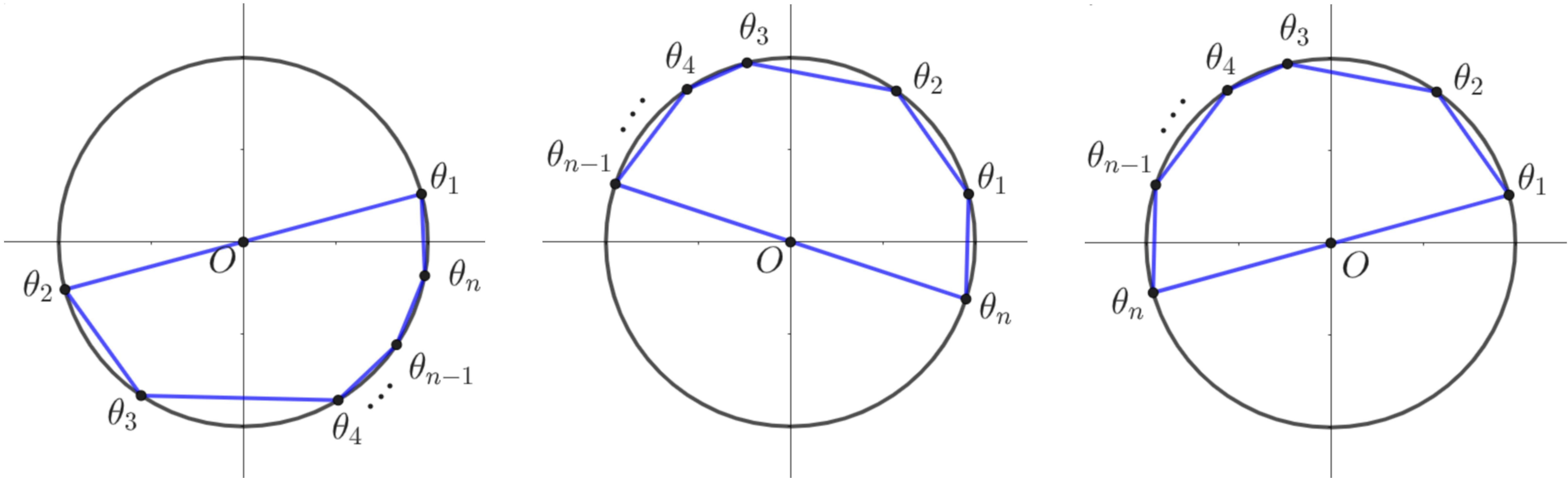}}
		\caption{For $n\geq2$, three of $n$ boundary situations corresponding to $\t_2-\t_1=\pi$, $\t_{n-1}-\t_{n}=\pi$ and $\t_{n}-\t_1=\pi$ are shown, respectively.    }
		\label{fig:schu=22}
	\end{figure*}
\end{proof}

Here we show the proof of Lemma \ref{le:minC1C2}.

\begin{proof}
(i) If $\cC_1=\cC_2$ then the equality in (\ref{eq:minc1c2}) holds. Next we assume that $\cC_1$ is a proper subset of $\cC_2$. Up to the subscript permutation, we can assume that  $\a_j=\b_{j}$ with $j=1,2,...,n$.  
As the proof of Lemma \ref{le: miniconv}, we consider
the polygon corresponding to $\text{conv}(\cC_1)$ and $\text{conv}(\cC_2)$, and denote them by $\cP_1$ and $\cP_2$, respectively. Since $\text{conv}(\cC_1)\subsetneq \text{conv}(\cC_2)$, $\cP_2$ is obtained by adding some points $\b_k$ with $k=n+1,...,m$ to $\cP_1$. 
Then the minimum distance between  $O$ and $\cP_1$ is not less than that between $O$ and $\cP_2$.  As above, the minimization in (\ref{eq:minc1c2}) can be transformed to the minimum distance between the original point $O$ and $\cP_1$ (resp. $\cP_2$). This completes the proof.

(ii) The assertion can be proved in the same way as (i).
\end{proof}

The proof of Lemma \ref{le:triepmax} is presented below.

\begin{proof}
	We show the case for $i=1$, others can be proved in a similar way.
For convenience of statement, one can further assume that $U=\sum_{i=1}^{d_{A_1}}(\proj{i})_{A_1}\otimes (U^{(i)})_{A_1^c}$, where $U^{(i)}$'s are diagonal unitaries. Without loss of generality,  we assume that the orthogonal projectors  $P_j=\sum_{i=M_{2j-1}}^{M_{2j}}\proj{i}$, where $M_1=1$, $M_{2m}=d_{A_1}$ and $M_{2j-2}+1=M_{2j-1}\leq M_{2j}$ for $j=1,2,...,m$. Since $U=\sum_{j=1}^m P_j \otimes U_j$, the unitaries satisfy $U_j=U^{(i)}$ with $M_{2j-1}\leq i\leq M_{2j}$.

 We show the derivation of the equality in (\ref{eq:max a1b1c1}).
Suppose  $\ket{\a_1,\a_1,...,\a_N}\in \ox_{k=1}^N \cH_{A_kR_k}$ is the critical state such that $K_{A_1:A_1^c}(U)=E_{A_1R_1:(A_1R_1)^c}(U(\ket{\a_1}\ox\ket{\a_2}\ox...\ox\ket{\a_N}))$, where $\ket{\a_1}=\sum_{k=1}^{d_{A_1}}\sqrt{p_k}\ket{k,a_k}_{A_1R_1}$, $p_k\geq 0$ and $\sum_kp_k=1$. From $U=\sum_{i=1}^{d_{A_1}}(\proj{i})_{A_1}\otimes (U^{(i)})_{A_1^c}$, we obtain that
\begin{eqnarray}
&&\!\!\!\!\!\!	\label{eq:E_{A:BC}}
K_{A_1:A_1^c}(U)=
E_{A_1R_1:(A_1R_1)^c}
\\
&&\!\!\!\!\!\!
 \times \bigg(
\sum_{i=1}^{d_{A_1}}\proj{i}\otimes U^{(i)} 
\Big(\sum_{k=1}^{d_{A_1}}\sqrt{p_k}\ket{k,a_k} \ox \ket{\a_2,...,\a_N}\Big)\bigg)
\nonumber\\
=&&\!\!\!\!\!\!
E_{A_1:(A_1R_1)^c}\bigg(U
\Big(\sum_{i=1}^{d_{A_1}}\sqrt{p_i}\ket{i}
\ox \ket{\a_2,...,\a_N}\Big)\bigg)
\nonumber\\
\leq 
&&\!\!\!\!\!\!
\max_{\substack{\ket{\ps_1'}\in\cH_{A_1}, \nonumber\\
\ket{\ps_k}\in\cH_{A_kR_k}, k\geq 2}}
E_{A_1:(A_1R_1)^c}
\big(U(\ket{\ps_1'}\ox \ket{\ps_2,...,\ps_N})\big)
\nonumber\\ &&\!\!\!\!\!\!
\leq  K_{A_1:A_1^c}(U).
\nonumber
\end{eqnarray}  
Hence, the ancilla $R_1$ can be removed. 

The equality in (\ref{eq:maxSqj}) is obtained by choosing $\ket{\ps_1'}=\sum_{k=1}^{d_{A_1}}\sqrt{p_k}\ket{k}$ with $p_k\geq 0$ and $\sum_k p_k=1$. From $U_j=U^{(i)}$ with $M_{2j-1}\leq i\leq M_{2j}$ for $j=1,2,...,m$, we have
\begin{eqnarray}
&&\!\!\!\!\!\!K_{A_1:A_1^c}(U)\\
=&&\!\!\!\!\!\!
\max_{\substack{p_k\geq 0, \sum_k p_k=1 \nonumber\\
\ket{\ps_k}\in\cH_{A_kR_k}, k\geq 2}}
E_{A_i:A_i^c}\bigg(\sum_{i=1}^{d_A}\proj{i}\otimes U^{(i)}
\nonumber\\ &&\hspace{2cm}
\times \Big( \sum_{k=1}^{d_{A_1}}\sqrt{p_k}\ket{k}\ox \ket{\ps_2,...,\ps_N}  \Big)\bigg)
\nonumber\\
=&&\!\!\!\!\!\!
\max_{\substack{q_j\geq 0, 
	\sum_{j=1}^m q_j=1\nonumber\\ \ket{\ps_k}\in\cH_{A_kR_k}, k\geq 2
}}
S\bigg(	\sum_{j=1}^m q_j U_j
\big(\ox_{k=2}^n \proj{\ps_k}\big) U_j^\dg
\bigg), \nonumber
\end{eqnarray}
where $q_j=\sum_{i=M_{2j-1}}^{M_{2j}}p_i$ and hence $q_j\geq 0$, $\sum_{j=1}^m q_j=1$. This complets the proof.
\end{proof}
\section{Analytical derivation of multi-qubit Schmidt rank two unitary operations}
\label{apx: n-qubit ep}
We present the proof of Proposition \ref{pro:ep3qubit} below.

\begin{proof}
	From \cite{cy13} $U$ can be controlled from every subsystem. Combined with Definition \ref{def:N-ep}, Proposition \ref{pro: ep max} and Lemma \ref{le:triepmax}, the entangling power of $U$ is given by 
	\begin{eqnarray}
	K_E(U)=&&\!\!\!\!\!\!
 \max_{\emptyset\subsetneq X\subsetneq \{A,B,C\}}   \Big(\max_{\ket{\a}\in\cH_{A},
			\ket{\b}\in\cH_{B},
			\ket{\g}\in\cH_{C}}
   \\ && \hspace{2cm}
		E_{X:X^c}(U(\ket{\a}\ox\ket{\b}\ox\ket{\g})) \Big)
	\nonumber	\\	
 =&& \!\!\!\!\!\!
 \max\big(K_{A:BC}(U), K_{B:AC}(U), K_{C:AB}(U)\big).
 \nonumber
	\end{eqnarray}
	
	First we consider the the maximal entanglement generation of $U$ under the bipartition $A:BC$. From Lemma \ref{le:triepmax}, we have
	\begin{eqnarray}
&&\!\!\!\!\!\!		K_{A:BC}(U)
=
\max_{p\in [0,1],\ket{\b}\in \cH_{B},
			\ket{\g}\in \cH_{C}}
   \\ &&
	\times	S(p\proj{\b,\g}
		+(1-p)U_1^\dg U_2\proj{\b,\g}U_2^\dg U_1), 
  \nonumber
	\end{eqnarray}
	where $\ket{\b}=(b_1,b_2)^T$, $\ket{\g}=(c_1,c_2)^T$ and $ b_s,c_t\geq 0$, as we can choose an appropriate local unitary $W_B\ox W_C$ acting on $\ket{\b,\g}$.
	Besides, there exists a  unitary $V$ acting on $\cH_{BC}$ whose first row is the transpose of $\ket{\b,\g}$. For convenience, we denote $\og_{0,0}, \og_{0,1}, \og_{1,0}, \og_{1,1}$ by $\og_1,\og_2,\og_3,\og_4$, respectively. We obtain that
	\begin{eqnarray}
	&&\!\!\!\!\!\!	K_{A:BC}(U)\\
=&&\!\!\!\!\!\!\max_{\substack{p\in [0,1],\ket{\b}=(b_1,b_2)^T,
\nonumber\\
\ket{\g}=(c_1,c_2)^T, b_s,c_t\geq 0}}
S\big(pV\proj{\b,\g}V^\dg
	\nonumber\\	&&\hspace{2.2cm}
  +(1-p)VU_1^\dg U_2\proj{\b,\g}U_2^\dg U_1V^\dg\big) 
	\nonumber	\\
=&&\hspace{-0.9cm}
\max_{\substack{p\in [0,1],\;\ket{\eta}=(x,\sqrt{1-x^2})^T,				\\x=\abs{\sum_{k=1}^{4}f_ke^{i(\og_k-\t_k)}}, f_k\geq 0, \sum_k f_k=1	}}
	\hspace{-0.7cm}	S\big(p\proj{0}
		+(1-p)\proj{\eta}\big) 
	\nonumber	\\
		=&&\hspace{-0.5cm} \max_{ f_k\geq 0, \;\sum_k f_k=1	}
		H\bigg(\frac{1+\b}{2},
		\frac{1-\b}{2}\bigg),
    \nonumber
	\end{eqnarray}
	where the second equality follows from $\{f_k\}_{k=1}^{4}=\{b_s^2c_t^2\}$ for $s=1,2$ and $t=1,2$. Hence $f_k\geq 0$ and $\sum_{k=1}^{4}f_k=1$.
	Without loss of generality, one can assume that  $0\leq\og_j-\t_j\leq\og_{j+1}-\t_{j+1}<2\pi$ for $1\leq j \leq 4$, and $\b=\abs{\sum_{k=1}^{4}
				f_ke^{i(\og_k-\t_k)}}$. 
	Using Lemma \ref{le: miniconv}, we have
\begin{eqnarray}
&&\hspace{-0.4cm}  K_{A:BC}(U)
\\
= &&\hspace{-0.4cm} 
\begin{cases}
1, 
\;\mbox{for} \;\;
0\leq\og_j-\og_{j-1}-(\t_{j}-\t_{j-1})\leq \pi,\;j=2,3,4,
\\\qquad \qquad \qquad
\mbox{and}\; \;
\og_{4}-\og_1-(\t_{4}-\t_1)\geq \pi,
\\
\max_{1\leq j\leq 4} H(\frac{1+\cos(\b_{j,k})}{2},
\frac{1-\cos(\b_{j,k})}{2}), \;\mbox{otherwise},
\end{cases}
\nonumber
\end{eqnarray}
where $\b_{j,k}=\frac{\og_j-\og_k-(\t_j-\t_k)}{2}$ with $k=j+1\mod 4$.

Next we consider $K_{B:AC}(U)$ and $K_{C:AB}(U)$.
First we rewrite $U$ in the form that $B$ is the control with $d_B$ terms.
By exchanging the system of $U_{ABC}$, we obtain that
\begin{eqnarray}
	&& \hspace{-0.4cm} U_{BAC}
 \\
=&&\hspace{-0.4cm} \sum_{k,t=0}^{1}\proj{k}_B\ox
	\big(	e^{i\t_{kt}}\proj{0,t}_{AC}	+e^{i\og_{kt}}\proj{1,t}_{AC}\big),
 \nonumber
\\
&&\hspace{-0.4cm} 	U_{CAB}
\\
=&&\hspace{-0.4cm} \sum_{k,t=0}^{1}\proj{t}_C\ox
	\big(
	e^{i\t_{kt}}\proj{0,k}_{AB}
	+
	e^{i\og_{kt}}\proj{1,k}_{AB}\big),
 \nonumber
\end{eqnarray}
Then by the same analysis as $K_{A:BC}(U)$, we  obtain $K_{B:AC}(U)$ and $K_{C:AB}(U)$. By taking maximization over the three bipartition, we
complete the proof.
\end{proof}

Here we show the Proof of Proposition \ref{pro:ep-nqubit}.

\begin{proof}
We analyze the entangling power of $U_{k}$ with respect to $k=n,n-1,2,1,0$, respectively. Here the unitary with $k=n$ or $n-1$ is analyzed in case (i); $k=2$ is analyzed in case (ii); $k=1$ or 0 is analyzed in case (iii).

(i) The Schmidt decomposition of a unitary whose singular number $k=n$ is given as 
\begin{eqnarray}
\label{def:Un}
U_{n}=I_2^{\ox n}+(e^{i\phi}-1)\proj{0}^{\ox n} \text{ with } \phi\in (0,2\pi).  
\end{eqnarray}
In order to analyze the entangling power of $U_{n}$, we should take the maximal entanglement generation across any bipartition by Definition \ref{def:N-ep} and Proposition \ref{pro: ep max}, that is,
\begin{eqnarray}
K_E(U_{n})
=&&\hspace{-0.4cm} \max_{\emptyset\subsetneq \L\subsetneq \{A_1R_{1},...,A_nR_{n}\}}
K_{\L:\L^c}(U_{n})
\\
=&&\hspace{-0.4cm} \max_{\emptyset\subsetneq \L\subsetneq \{A_1R_{1},...,A_nR_{n}\}}
	   \Big(
\max_{\ket{\ps_i}\in\cH_{A_iR_{i}}, i=1,2,...,n}
\nonumber\\&& \hspace{-0.4cm} 
\times E_{\L:\L^c}(U_{n}(\ket{\ps_1}_{A_1R_{1}}\ox...\ox\ket{\ps_n}_{A_n R_{n}})) \Big),
\nonumber
\end{eqnarray}
Across any bipartition $\L:\L^c$, the unitary can be treated as a bipartite unitary. It has been shown that either party of this unitary can be chosen as control in \cite{cy13}.   From Lemma \ref{le:m:nbipartition}, the ancilla of each system can be removed.   Note that $U_n$ is symmetric on all systems $A_1,A_2,...,A_n$, that is, the expression of $U_{n}$ is the same up to the permutation of systems.  Due to the symmetry of $U_{n}$, it suffices to consider $[\frac{n}{2}]$ types of  bipartition, i.e.  $\L_t=A_1...A_t$ and $\L^c_t=A_{t+1}...A_n$ for $t=1,2,...,[\frac{n}{2}]$.
With this bipartition, we write $U_{n}$ in the form of two-term control as follows
\begin{eqnarray}
U_{n}
=&&\hspace{-0.4cm} \proj{0}^{\ox t} _{\L_t}
\ox (I_2^{\ox (n-t)}+(e^{i\t}-1)\proj{0}^{\ox (n-t)})_{\L_t^c}
\nonumber\\
+&&\hspace{-0.4cm} (I_2^{\ox t}-\proj{0}^{\ox t})_{\L_t}\ox (I_2^{\ox (n-t)})_{\L_t^c}
\nonumber\\
:=&&\hspace{-0.4cm} (P_1^{(1)})_{\L_t}\ox (U_1^{(1)})_{\L_t^c}+(P_2^{(1)})_{\L_t}\ox (U_2^{(1)})_{\L_t^c}.
\nonumber
\end{eqnarray}
  If we exchange the systems $\L_t$ and $\L^c_t$ of this unitary, then the entanglement generation $K_{\L_t:\L_t^c}(U_{n})$ remains the same.
Therefore, we analyze the entanglement generation of $U_{n}$ by choosing either $\L_t$ or $\L_t^c$ as the control and will obtain the same result. 
Using Lemma \ref{pro:schu=2}, we choose $\L_t$ as the control and  obtain that for $t=1,2,...,[\frac{n}{2}]$,
\begin{eqnarray}
K_{\L_t:\L_t^c}(U_{n})
=
H(\frac{1+\cos(\phi/2)}{2},
\frac{1-\cos(\phi/2)}{2}), 
\nonumber
\end{eqnarray}
for $\phi\in (0,2\pi)$.
Hence, the entangling power of $U_{n}$ is give as follows. For $\phi\in (0,2\pi)$, 
\begin{eqnarray}
\label{eq:keUk=n}
K_E(U_{n})=&&\hspace{-0.4cm}
\max_{t\in \{1,2,...,n-1\}}
K_{\L_t:\L_t^c}(U_{n})
\\
=&&\hspace{-0.4cm}
H(\frac{1+\cos(\phi/2)}{2},
\frac{1-\cos(\phi/2)}{2}).
\nonumber
\end{eqnarray}

 When a unitary with the singular number $k=n-1$, its Schmidt decomposition is given as 
 \begin{eqnarray}
\label{def:Un-1}
\hspace{-0.4cm} 
U_{n-1}=I_2^{\ox n}+\proj{0}^{\ox (n-1)}\ox \diag(e^{i\t}-1,e^{i\phi}-1) 
 \end{eqnarray}
with $\t,\phi\in (0,2\pi)$ and $\t\neq \phi$. 
The unitary is symmetric except for the last system $A_n$. By some observations, it can be obtained that when the system $A_n$ is not performed as one system of the control under a proper bipartition, the unitary $U_{n-1}$ can always be written in the form of two-term control. 
Since the systems $A_1,A_2,...,A_{n-1}$ are symmetric, it suffices to analyze the entangling power of this unitary under the bipartition $\L_t=A_1...A_t$ and $\L^c_t=A_{t+1}...A_n$ for $t=1,2,...,n-1$ and choose $\L_t$ as the control. Then $U_{n-1}$ can be given by
\begin{eqnarray}
U_{n-1}
=&&\hspace{-0.4cm}\proj{0}^{\ox t} _{\L_t}
\ox (I_2^{\ox (n-t)}
+(e^{i\t}-1)\proj{0}^{\ox (n-t)}
\nonumber\\ && \hspace{-0.4cm}
+(e^{i\phi}-1)\proj{0}^{\ox (n-t-1)}\ox\proj{1})_{\L_t^c}
\nonumber\\
+&&\hspace{-0.4cm}(I_2^{\ox t}-\proj{0}^{\ox t})_{\L_t}\ox (I_2^{\ox (n-t)})_{\L_t^c}
\nonumber
\\
:=&&\hspace{-0.4cm}(P_1^{(2)})_{\L_t}\ox (U_1^{(2)})_{\L_t^c}+(P_2^{(2)})_{\L_t}\ox (U_2^{(2)})_{\L_t^c}.
\nonumber
\end{eqnarray}
It is worth mentioning that choosing $\L_t$ as the control to analyze the entangling power is
without loss of generality, as the entangling generation remains the same when we exchange the two systems, i.e., $K_{\L_t:\L_t^c}(U_{\L_t\L_t^c})=K_{\L_t:\L_t^c}(U_{\L_t^c\L_t})$. We assume that $\l_1=0, \l_2=\min\{\t,\phi\}$ and $\l_3=\max\{\t,\phi\}$. Then we obtain the entangling power of $U_{n-1}$ as follows
\begin{eqnarray}
\label{eq:keUk=n-1}
&&\hspace{-0.4cm}K_E(U_{n-1})=\hspace{-0.4cm}\max_{t\in \{1,2,...,n-1\}}
K_{\L_t:\L_t^c}(U_{n-1})
\nonumber
\\
=&& \hspace{-0.4cm}
\begin{cases}
H(\frac{1}{2},\frac{1}{2})=1, 
\mbox{for} \;
0<\l_j-\l_{j-1}\leq \pi,\;j=2,3,
\l_{3}\geq \pi,
 \\
\max_{1\leq j \leq 3} H(\frac{1+\cos(\b)}{2},
\frac{1-\cos(\b)}{2}),\;\mbox{otherwise},
\end{cases}
\nonumber
\end{eqnarray}
where $\b=\frac{\l_j-\l_{j+1\mod 3}}{2}$.

(ii) A unitary with the singular number $k=2$ is given as, for $\b_j\in (0,2\pi)$, 
\begin{eqnarray}
\label{def:Uk=2}
U_{2}=\proj{0}\ox I_2^{\ox (n-1)}+\proj{1} \ox \big(\ox_{j=2}^n \diag(1,e^{i\b_j})\big).\nonumber\\  
\end{eqnarray} 
The systems of $U_{2}$ are not symmetric. In order to obtain the entangling power of $U_{2}$, all kinds of bipartition of it should be taken into account. 

First, we show the expression of $U_{2}$ a kind of bipartition, i.e.  $\L_t=A_1...A_t$ and $\L^c_t=A_{t+1}...A_n$ for $t=1,2,...,n-1$.  Without loss of generality, we choose the control as $\L_t$. Then $U_{2}$ is controlled with two terms as follows
\begin{eqnarray}
U_{2}=&& \hspace{-0.4cm}
\sum_{j_2,j_3,...,j_t=0}^1
\proj{0,j_2,j_3,...,j_t} \ox I_2^{\ox (n-t)}
\\
+&&\hspace{-0.4cm}\sum_{j_2,j_3,...,j_t=0}^1
e^{i(\sum_{s=2}^t\b_s\d_{1,j_s})}\proj{1,j_2,j_3,...,j_t} 
\nonumber\\&& \hspace{4cm}
\ox \big( \ox_{j=t+1}^n \diag(1,e^{i\b_j}) \big),
\nonumber
\end{eqnarray}
where $\d_{x,y}=1(\text{resp. 0})$ for $x=y$ (resp. $x\neq y$).
Performing a local unitary $V\in \cU_{A_1A_2,...,A_t}$ on $U_{2}$, we obtain a unitary
\begin{eqnarray}
\label{eq:U'k=2}
U_{2}'=&&\hspace{-0.4cm}(V\ox I_2^{\ox(n-t)})U_{2}
\\
=&&\hspace{-0.4cm}\sum_{j_2,j_3,...,j_t=0}^1
\proj{0,j_2,j_3,...,j_t} \ox I_2^{\ox (n-t)}
\nonumber\\
+&&\hspace{-0.4cm}\sum_{j_2,j_3,...,j_t=0}^1
\proj{1,j_2,j_3,...,j_t} 
\nonumber\\ &&\hspace{4cm}
\ox \big( \ox_{l=t+1}^n \diag(1,e^{i\b_l}) \big)
\nonumber\\
:=&& \hspace{-0.4cm} (P_1^{(3)})_{\L_t}\ox (U_1^{(3)})_{\L^c_t}+(P_2^{(3)})_{\L_t} \ox (U_2^{(3)})_{\L^c_t}.
\nonumber
\end{eqnarray}
By the locally unitary  invariance of von Neumann entropy, $K_{\L_t:\L_t^C}(U_{2})=K_{\L_t:\L_t^C}(U_{2}')$.
Suppose $\cA_t$ with $\abs{\cA_t}=2^{n-t}$ is the set consisting of the phases of all diagonal elements of the unitary $U_2^{(3)}=\ox_{l=t+1}^n \diag(1,e^{i\b_l})$. We rearrange the elements of $\cA_t$ in the ascending order and denote the $j$-th smallest element in this set by $\t_j$, for $j=1,2,...,2^{n-t}$.
That is, 
\begin{eqnarray}
\label{def:cA_t}
\cA_t=&& \hspace{-0.4cm}\left\{ \exp\bigg[i\big(\sum_{l=t+1}^n q_l\b_l\mod 2\pi\big)\bigg] \bigg|q_l=0,1\right\}
\\
=&& \hspace{-0.4cm}\{e^{i\t_j} | j=1,2,...,2^{n-t}, \t_p\leq \t_q
, 1\leq p<q\leq 2^{n-t}\}.
\nonumber
\end{eqnarray}
By the same analysis as in Lemma \ref{pro:schu=2}, the entanglement generation of $U_{2}$ under this bipartition is 
\begin{eqnarray}
\label{eq:K_Lt:Ltc}
&&\hspace{-0.4cm}K_{\L_t:\L_t^c}(U_{2})
\\
=&&\hspace{-0.4cm}
\begin{cases}
1, 
\;\mbox{for} \;
0\leq\t_j-\t_{j-1}\leq \pi,
\t_{2^{n-t}}-\t_1\geq \pi,\\
\max_{1\leq j \leq 2^{n-t}} H(\frac{1+\cos(\b)}{2},
\frac{1-\cos(\b)}{2}, 
\;\mbox{otherwise},
\end{cases}
\nonumber
\end{eqnarray}
where $j=2,3,...,2^{n-t}$, and $\b=\frac{\t_j-\t_k}{2}$  with  $k=(j+1)\mod 2^{n-t}$.
Next we compare $K_{\L_t:\L_t^c}(U'_{k=2})$ over the bipartition $\L_t:\L_t^c$ with $t=1,2,...,[\frac{n}{2}]$. We claim that $K_{\L_t:\L_t^c}(U'_{k=2})$ is non-increasing with respect to $t$. In fact, one can obtain that
\begin{eqnarray}
\label{eq:K_Lt}
K_{\L_t:\L_t^c}(U'_{k=2})\geq K_{\L_{t+1}:\L_{t+1}^c}(U'_{k=2}).  
\end{eqnarray}
 From  (\ref{eq:U'k=2}), $U_2^{(3)}=\ox _{l=t+1}^n\diag(1,e^{i\b_l})$ (resp. $\ox _{l=t+2}^n\diag(1,e^{i\b_l})$) by the bipartition $\L_t:\L_t^c$ (resp. $\L_{t+1}:\L_{t+1}^c$). Hence the set $\cA_t$ in (\ref{def:cA_t}) satisfies $\cA_{t+1}\subset \cA_t$. 
We denote the minimum convex sum of $\cA_{t}$ by $\min \text{conv}(\cA_t)$. Then  we have 
\begin{eqnarray}
\label{eq: KLt+1<=KLt}
&& \hspace{-0.4cm}
K_{\L_{t+1}:\L_{t+1}^c}(U'_{k=2})
\\
=&& \hspace{-0.4cm}
H(\frac{1- \min \text{conv}(\cA_{t+1})}{2}, \frac{1+ \min \text{conv}(\cA_{t+1})}{2} )
\nonumber\\
\leq && \hspace{-0.4cm}
H(\frac{1- \min \text{conv}(\cA_{t})}{2}, \frac{1+ \min \text{conv}(\cA_{t})}{2} )
\nonumber \\
= &&\hspace{-0.4cm} K_{\L_t:\L_t^c}(U'_{k=2}),   \nonumber
\end{eqnarray}
where the first and third equality follows from (\ref{eq: schu=2 keu}), and the second equality is derived by $\cA_{t+1}\subset \cA_t$ and  Lemma \ref{le:minC1C2}. By now we have proven (\ref{eq:K_Lt}). 
From (\ref{eq:K_Lt}), we obtain that
\begin{eqnarray}
\label{eq:maxKA1}
\max_{t\in \{1,2,...,[\frac{n}{2}]\}}
K_{\L_t:\L_t^c}(U_{2})=K_{A_1:A_1^c}(U_{2}). 
\end{eqnarray}
Therefore, in order to consider the maximal entanglement generation under the bipartition $\L_t:\L_t^c$, we can only consider the bipartition $A_1:A_1^c$. Let $\cA_1=\{ (\sum_{l=2}^n q_l\b_l)\mod 2\pi|q_l=0,1\}=\{\t_j | j=1,2,...,2^{n-1}\;\text{and}\; \t_p\leq \t_q\;\text{for}\; 1\leq p<q\leq 2^{n-1}\}$. 
By choosing $t=1$ in (\ref{eq:K_Lt:Ltc}), one has the entanglement generation over $A_1:A_1^c$ as 
\begin{eqnarray}
\label{eq:KE3}
&&\hspace{-0.4cm}K_{A_1:A_1^c}(U_{2})
\\
=&&\hspace{-0.4cm}
\begin{cases}
1, 
\;\mbox{for} \;\;
0\leq\t_j-\t_{j-1}\leq \pi,
,
\t_{2^{n-1}}\geq \pi,
\nonumber\\
\max_{1\leq j \leq 2^{n-1}} H(\frac{1+\cos(\b)}{2},
\frac{1-\cos(\b)}{2}),\; \mbox{otherwise},
\end{cases}
\end{eqnarray}
where $j=2,3,...,2^{n-1}$, and $\b=\frac{\t_j-\t_k}{2}$ with $k=j+1\mod 2^{n-1}$.

Next we consider other bipartition $\Omega:\Omega^c$, where  $\emptyset \subsetneq \Omega \subsetneq \{A_1,A_2,...,A_n\}$. As we have shown in (\ref{eq:maxKA1}), the fewer systems working as the control under the bipartition $\L_t:\L_t^c$ there are, the larger the entangling power will be. Hence the maximal entanglement generation is reached when a single system works as the control, across any bipartition $\Omega:\Omega^c$.
In summary, the entangling power of $U_{2}$ in (\ref{def:Uk=2}) is obtained by taking the maximization over the following $n$ types of bipartition, that is,
\begin{eqnarray}
\label{eq:K_E1}
\max_{\emptyset\subsetneq \Omega\subsetneq \{A_1,...,A_n\}}
K_{\Omega:\Omega^c}(U_{2})
=
\max_{t\in \{1,2,...,n\}}
K_{A_t:A_t^c}(U_{2}).
\nonumber
\\
\end{eqnarray}
We write $U_{2}$ in (\ref{def:Uk=2}) in the form that the system $A_t$ is the control with $t=2,3,...,n$, i.e.
\begin{eqnarray}
U_{2}
=&&\hspace{-0.4cm}\proj{0}_{A_t}
\ox \Big(\proj{0}_{A_1}\ox (I_2^{\ox (n-2)})_{(A_1A_t)^c}\nonumber\\
&&+\proj{1}_{A_1}\ox (\ox_{j\neq t}\diag(1,e^{i\b_j}))_{(A_1A_t)^c}\Big)
\nonumber\\
+&&\hspace{-0.4cm}\proj{1}_{A_t}
\ox \Big(\proj{0}_{A_1}\ox (I_2^{\ox (n-2)})_{(A_1A_t)^c}\nonumber\\
&&+e^{i\b_t}\proj{1}_{A_1}\ox (\ox_{j\neq t}\diag(1,e^{i\b_j}))_{(A_1A_t)^c}\Big),\nonumber
\end{eqnarray}
which is locally unitarily equivalent to 
\begin{eqnarray}
\label{eq:U''k=2}
U_{2}''
=&&\hspace{-0.4cm}\proj{0}_{A_t}
\ox (I_2^{\ox (n-1)})_{A_t^c}
\nonumber
\\
+&&\hspace{-0.4cm}\proj{1}_{A_t}
\ox \Big(\proj{0}_{A_1}\ox (I_2^{\ox (n-2)})_{(A_1A_t)^c}\nonumber\\
&&\hspace{1cm}+
e^{i\b_t}\proj{1}_{A_1}\ox (I_2^{\ox (n-2)})_{(A_1A_t)^c}\Big) 
\nonumber
\end{eqnarray}
with respect to the bipartition $A_t:A_t^c$.
From Lemma \ref{pro:schu=2} and (\ref{eq:U''k=2}), we obtain that for $t=2,3,...,n$,
\begin{eqnarray}
\label{eq:KE2}
&& \hspace{-0.4cm}
K_{A_t:A_t^c}(U_{2})=K_{A_t:A_t^c}(U''_{2})
\\
=&& \hspace{-0.4cm} H(\frac{1+\cos(\b_t/2)}{2},
\frac{1-\cos(\b_t/2)}{2}), 
\nonumber
\end{eqnarray}
for $\b_t\in (0,2\pi)$.
We have shown the case $t=1$ by (\ref{eq:KE3}) and $t>1$ by (\ref{eq:KE2}).
Then we obtain the entangling power of the unitary $U_{2}$. Hence the equality (\ref{eq:ep k=2}) holds for 
$\cA_1=\{e^{i\t_j} | j=1,2,...,2^{n-t}\;\text{and}\; \t_p\leq \t_q\;\text{for}\; 1\leq p<q\leq 2^{n-t}\}$.

(iii) A $n$-qubit unitary with singular number $k=1$ is given by 
\begin{eqnarray}
\label{def:U1}
U_{1}=
i\cos\a \proj{0}\ox \s_3^{\ox (n-1)}+\diag(\sin\a,1)\ox I_2^{\ox (n-1)}   \nonumber
\\
\end{eqnarray}
with  $\a\in (0,\frac{\pi}{2})\cup (\frac{\pi}{2},\pi)$.  
 The unitary is symmetric except for the first system. We can merely analyze the entanglement generation under the bipartition $\L_t=A_1...A_t$ and $\L_t^c=A_{t+1}...A_n$ for $t=1,2,...,n-1$. Under this bipartition, the unitary is given as 
\begin{eqnarray}
\label{eq:U_k=1}
 U_{1}=&&\!\!\!\!\!\!
\bigg[
\sum_{j_2...j_n} \Big(\cos\a+(-1)^{\sum_{s=2}^n\d_{1,j_s}}i\sin\a \Big)
\\ && \hspace{-0.4cm}
\proj{0,j_2,...,j_t}+\proj{1,j_2,...,j_t} 
\bigg]_{A_1...A_t}
\nonumber
\\&& \!\!\!\!\!\!
\ox \big[ \proj{j_{t+1},...,j_n}  \big]
 _{A_{t+1}...A_n}. \nonumber
\end{eqnarray}
Note that $\sum_{s=2}^n\d_{1,j_s}$ is either odd or even. Based on  its value, we divide the set of POVMs of systems $A_{t+1}...A_n$ into two subsets. That is, 
\begin{eqnarray}
\label{eq:P1}
&&\hspace{-0.4cm}P_1^{(4)}=\sum_{g_{t+1}...g_n} \proj{g_{t+1}...g_n},\;\text{for odd}\sum_{s=t+1}^n\d_{1,g_s},
\nonumber\\
\\
\label{eq:P2}
&&\hspace{-0.4cm}P_2^{(4)}=\sum_{h_{t+1}...h_n} \proj{h_{t+1}...h_n},\;\text{for even}\sum_{s=t+1}^n\d_{1,h_s}.
\nonumber
\\
\end{eqnarray}
Obviously, $P_1^{(4)}+P_2^{(4)}=I_2^{\ox(n-t)}$ and $P_1^{(4)}$ is orthogonal to $P_2^{(4)}$. 
From (\ref{eq:U_k=1})-(\ref{eq:P2}), one can obtain that 
\begin{eqnarray}
\hspace{-0.4cm}
U_{1}=(U_1^{(4)})_{\L_t}\ox (P_1^{(4)})_{\L_t^c}+ (U_2^{(4)})_{\L_t}\ox (P_2^{(4)})_{\L_t^c},
\end{eqnarray}
where the unitaries
\begin{eqnarray}
\label{eq:U_1^4}
U_1^{(4)}=&&\hspace{-0.4cm}
\sum_{j_2...j_t}\Big(\cos\a-(-1)^{\sum_{s=2}^t\d_{1,j_s}}i\sin\a\Big)
\\ &&\hspace{-0.4cm}
\times\proj{0,j_2,...,j_t}+\proj{1,j_2,...,j_t}, 
\nonumber
\\
\label{eq:U_2^4}
U_2^{(4)}=&&\hspace{-0.4cm}
\sum_{j_2...j_t}\Big(\cos\a+(-1)^{\sum_{s=2}^t\d_{1,j_s}}i\sin\a\Big)
\\
&&\hspace{-0.4cm}
\times\proj{0,j_2,...,j_t}+\proj{1,j_2,...,j_t}. 
\nonumber
\end{eqnarray}
Obviously, $U_{1}$ is locally unitarily equivalent to 
\begin{eqnarray}
U_{1}'
=(I_2^{\ox t})_{\L_t}\ox (P_1^{(4)})_{\L_t^c}+ ((U_1^{(4)})^\dg U_2^{(4)})_{\L_t}\ox (P_2^{(4)})_{\L_t^c}.
\nonumber
\\
\end{eqnarray}
From (\ref{eq:U_1^4}) and (\ref{eq:U_2^4}), the diagonal elements of $(U_1^{(4)})^\dg U_2^{(4)}$ are $1,\exp[\pm2\a i]$.
Let the set $\cB=\{0, \pm2\a \mod 2\pi\}
=\{\l_j|j=1,2,3,\;\l_1\leq \l_2\leq \l_3 \}$.
From Lemma \ref{pro:schu=2}, we obtain the entangling power of $U_{1}$ as follows
\begin{eqnarray}
\label{eq:keUk=1}
&&\hspace{-0.4cm}K_E(U_{1})
=\max_{t\in \{1,2,...,n-1\}}
K_{\L_t:\L_t^c}(U_{1})
\nonumber
\\
=&&\hspace{-0.4cm}
\begin{cases}
1, 
\;\mbox{for} \;
0\leq \l_j-\l_{j-1}\leq \pi,\;j=2,3,
\l_{3}\geq \pi,
 \\
\max_{1\leq j\leq 3} H(\frac{1+\cos(\b)}{2},
\frac{1-\cos(\b)}{2}),\;\mbox{otherwise},
\end{cases}
\end{eqnarray}
where $\b=(\l_j-\l_{j+1\mod 3})/2$.

 When the singular number $k=0$, the unitary is written as 
 \begin{eqnarray}
 \label{def:U0}
\hspace{-0.8cm}U_{0}=&&\hspace{-0.4cm}\frac{\l_j-\l_{j+1\mod 3}}{2}i\diag(\sin\a,\sin\b) \ox \s_3^{(n-1)}
\\+&&\hspace{-0.4cm}\frac{\l_j-\l_{j+1\mod 3}}{2}\diag(\cos\a,\cos\b)\ox I_2^{\ox (n-1)}   \nonumber
 \end{eqnarray} 
  with  
$\a,\b\in (0,\frac{\pi}{2})\cup (\frac{\pi}{2},\pi)$.
 By similar analysis as for $k=1$, the unitary is written in the form of two-term control, across the bipartition $\L_t=A_1...A_t$ and $\L_t^c=A_{t+1}...A_n$ as follows 
\begin{eqnarray}
\hspace{-0.4cm}
U_{0}=
(U_1^{(5)})_{\L_t}\ox (P_1^{(5)})_{\L_t^c}+ (U_2^{(5)})_{\L_t}\ox (P_2^{(5)})_{\L_t^c},
\end{eqnarray}
where $P_t^{(5)}=P_t^{(4)}$ in (\ref{eq:P1}) and (\ref{eq:P2}) with $t=1,2$, and
\begin{eqnarray}
\label{eq:U_1^5}
U_1^{(5)}
=&&\hspace{-0.4cm}\sum_{j_2...j_t}\Big(\cos\a-(-1)^{\sum_{s=2}^t\d_{1,j_s}}i\sin\a\Big)
\\
&& \qquad \times \proj{0,j_2,...,j_t}
\nonumber\\
+&&\hspace{-0.4cm}\Big(\cos\b-(-1)^{\sum_{s=2}^t\d_{1,j_s}}i\sin\b \Big)
\nonumber\\
&& \qquad \times
\proj{1,j_2,...,j_t}, 
\nonumber
\\
\label{eq:U_2^5}
U_2^{(5)}=&&\hspace{-0.4cm}\sum_{j_2...j_t}\Big(\cos\a+(-1)^{\sum_{s=2}^t\d_{1,j_s}}i\sin\a\Big)
\\
&& \qquad \times
\proj{0,j_2,...,j_t}
\nonumber\\
+&&\hspace{-0.4cm}\Big(\cos\b+(-1)^{\sum_{s=2}^t\d_{1,j_s}}i\sin\b \Big)
\nonumber\\
&& \qquad \times
\proj{1,j_2,...,j_t}.
\nonumber
\end{eqnarray}
The unitary $U_{0}$ is locally equivalent to 
\begin{eqnarray}
U_{0}=
(I_2^{\ox t})_{\L_t}\ox (P_1^{(5)})_{\L_t^c}+ ((U_1^{(5)})^\dg U_2^{(5)})_{\L_t}\ox (P_2^{(5)})_{\L_t^c}.
\nonumber
\end{eqnarray}
The diagonal elements of $(U_1^{(5)})^\dg U_2^{(5)}$ are $\exp[\pm 2\a i], \exp[\pm 2\b i]$.
Let $\cC=\{(\pm2\a)\mod 2\pi, (\pm2\b)\mod 2\pi\}
=\{\l_j|j=1,2,3,4,\;\l_1\leq \l_2\leq\l_3\leq\l_4 \}$.
From Lemma \ref{pro:schu=2}, we obtain the entangling power of $U_{0}$ shown in (\ref{eq:ep k=1,0}). 
This completes the proof.
\end{proof}
\bibliographystyle{unsrt}
\bibliography{ep}
\end{document}